\pdfoutput=1
\PassOptionsToPackage{hyphens}{url}
\documentclass[a4paper,fleqn]{cas-sc}

\usepackage[numbers]{natbib}

\usepackage{siunitx}
\usepackage{amsmath,amssymb}
\usepackage{etoc}

\usepackage{float}
\usepackage[fontsize=10pt]{fontsize}
\definecolor{detblue}{HTML}{1F77B4}
\definecolor{bayorange}{HTML}{FF7F0E}
\definecolor{failred}{HTML}{D62728}

\graphicspath{{Figures/}}

\begin{document}
\let\WriteBookmarks\relax
\def\floatpagepagefraction{1}
\def\textpagefraction{.001}

\shorttitle{Dosimetric equivalence of prostatic DVH indices from automatic contours, with predictive value for IPSS after LDR-BT}

\shortauthors{L.-B. St-Cyr et~al.}

\title [mode = title]{Dosimetric equivalence of prostatic DVH indices from automatic contours, with incremental predictive value for IPSS after LDR brachytherapy}

\author[1,4]{Louis-Bernard St-Cyr}[type=editor]
\credit{Conceptualization, Methodology, Software,
Formal analysis, Investigation, Data curation, Visualization,
Writing -- original draft.}

\affiliation[1]{organization={Université Laval, Département de physique, génie physique et d'optique},
    city={Québec, QC},
    country={Canada}}
\affiliation[2]{organization={ Radiation Oncology Service, Department of Specialized Medicine, Centre Intégré de Cancérologie (CIC), Hôpital de l’Enfant-Jésus, Centre Hospitalier Universitaire (CHU) de Québec-Université Laval},
    city={Québec, QC},
    country={Canada}}
\affiliation[3]{organization={Service of Radio-oncology, CHU de Québec-Université Laval},
city={Québec, QC},
country={Canada}}
\affiliation[4]{organization={ Oncology division, Research Center of the CHU de Québec-Université Laval},
city={Québec, QC},
country={Canada}}
\affiliation[5]{organization={Université Laval, Faculté de médecine},
    city={Québec, QC},
    country={Canada}}

\author[5]{Anne Saint-Laurent}
\credit{Data curation, Investigation}
\author[1,4]{José Angel Lesteiro-Tejeda}
\credit{Resources, Data curation, Writing -- review \& editing}
\author[3]{Sylviane Aubin}
\credit{Data curation, Writing -- review \& editing}
\author[3]{Marie-Claude Lavallée}
\credit{Writing -- review \& editing}
\author[1,3,4]{Luc Beaulieu}
\credit{Writing -- review \& editing}
\author[2,4]{Éric Vigneault}
\credit{Resources, Writing -- review \& editing}
\author[2,4]{André-Guy Martin}
\credit{Resources, Writing -- review \& editing}
\author[2,3]{François-Olivier Fabi}\fnmark[1]
\credit{Conceptualization, Supervision,
Project administration, Writing -- review \& editing}
\author[1,3,4]{Louis Archambault}
\credit{Conceptualization, Supervision,
Project administration, Funding acquisition,
Writing -- review \& editing}
\fnmark[1]
\cormark[1]
\ead{Louis.Archambault@phy.ulaval.ca}

\cortext[cor1]{Corresponding author}
\fntext[1]{Co-senior author / These authors contributed equally to the supervision/direction of this work.}

\begin{abstract}
\noindent\textbf{\textit{Background and purpose:}} Dose Volume Histogram (DVH) indices remain the main dose-effect metrics for toxicity prediction, but they depend on how contours were made. Inter-observer variability (IOV) is unavoidable and clinically accepted, so the question for automatic segmentation addresses equivalence: do automatic contours produce DVH errors comparable to human IOV, and can these indices predict patient-reported toxicity?

\noindent\textbf{\textit{Materials and methods:}} In 429 patients treated with iodine-125 LDR-BT monotherapy, indices from expert manual delineation were compared with a deterministic and a Bayesian nnUNet by two one-sided tests against margins set from CT contouring IOV. Logistic regression gave patient-level thresholds at 90\,\% probability of equivalence. In 380 patients, eleven DVH indices were added to a clinical baseline predicting change in International Prostate Symptom Score (IPSS) at six horizons over 5 years, with corrected $t$-tests.

\noindent\textbf{\textit{Results:}}
All cohort-level comparisons of DVH indices were declared equivalent, with no interval consuming more than 44\,\% of its equivalence margin. Individual agreement was weaker with equivalence rates from 55.9\,\% to 89.7\,\%, depending on the DVH index and segmentation model. Thresholds ranged from 0.864 to 0.958 Dice. No DVH block improved IPSS prediction at any horizon. The largest improvement declared by bootstrap intervals was 0.12 IPSS points, well below the gain equivalent to the MCID.

\noindent\textbf{\textit{Conclusion:}} Automatic contours matched expert dosimetry within human IOV on average, but individual equivalence depends on prostate volume and contour quality, defining a commissioning criterion for a segmentation model. Our DVH indices panel provides no significant predictive power for IPSS, irrespective of segmentation source and horizon in this cohort.
\end{abstract}

\begin{highlights}
\item Automatic prostate contours matched expert DVH indices within human observer noise.
\item Patient-level Dice thresholds for dosimetric equivalence depend on prostate volume.
\item DVH indices added no predictive value for IPSS change at six horizons over 5 years.
\item The best improvement in IPSS change predictive value is well below the MCID.
\end{highlights}


\maketitle
\etocdepthtag.toc{mainmatter}

\section{Introduction}\label{sec:intro}

Low-dose-rate brachytherapy (LDR-BT) is an effective monotherapy option for
low-risk and favourable intermediate-risk prostate cancer~\cite{mohlerProstateCancerVersion2012}, with high long-term
biochemical control. Urinary toxicity nonetheless remains the
dominant morbidity~\cite{keyesPredictiveFactorsAcute2009a,tanimotoPredictiveFactorsAcute2013}, and is tracked through patient-reported outcome surveys such as
the International Prostate Symptom Score (IPSS)~\cite{gregoireValidationFrenchAdaptation1996}. Dose distribution and dose-volume indices are computed on a post-implant CT, where organ delineation can represent a real reproducibility issue~\cite{crookInterobserverVariationPostimplant2002b,debrabandereProstatePostimplantDosimetry2012,lindsaySystematicStudyImaging2003}.
 
Automatic segmentation is often evaluated by geometric agreement with expert
contours, through overlap and surface metrics or clinical acceptability ratings~\cite{heilemannClinicalImplementationEvaluation2023,duanEvaluatingClinicalAcceptability2022,duanIncrementalRetrainingClinical2023},
not taking into consideration the dose distribution~\cite{vanaalstDosebasedEvaluationDelineation2026}.
When that link is made, the questions asked are whether the dose-volume
histogram (DVH) indices derived from automatic contours differ significantly from
the expert, and whether they remain acceptable for treatment planning~\cite{kawulaDosimetricImpactDeep2022,arjmandiEvaluatingDosimetricImpact2025,sritharanDosimetricComparisonAutomatically2022,hoqueClinicalUseCommercial2023}. Dosimetric equivalence is rarely studied, although it is what matters: an automatic contour can be considered dosimetrically equivalent to the expert when the dosimetric
error it produces is similar to the error two experts already produce
between themselves~\cite{debrabandereProstatePostimplantDosimetry2012,kirisitsReviewClinicalBrachytherapy2014}. Failing to detect a significant dosimetric difference does not imply equivalence~\cite{lakensEquivalenceTestsPractical2017}. Whether the segmentation source alters the predictive value of DVH indices for IPSS scores evolution has not been examined, the closest precedent being on planning outcomes rather than toxicity~\cite{lebaoEvaluatingRelationshipContouring2024}.
 
We address both the equivalence and predictive questions on a single-institution cohort (CHU de Québec-Université Laval) treated with LDR-BT as
monotherapy, by comparing expert manual contours with deterministic and Bayesian
nnUNet~\cite{isenseeNnUNetSelfconfiguringMethod2021a} against equivalence margins declared in advance from inter-observer variability (IOV). We also test whether DVH indices improve the prediction of
IPSS scores evolution over a clinical baseline at six post-implant
horizons.

\section{Materials and methods}\label{sec:Methods}

\subsection{Cohort and endpoints}\label{subsec:cohort_endpoints}

We retrospectively analysed patients treated between 2004 and 2024 at a single institution with LDR-BT as monotherapy for prostate cancer, with 144 or 145~Gy
prescriptions using iodine-125 seeds. In our cohort, 448 patients had sufficient imaging data with available records to be analyzed. 18 of those patients were
excluded because their images had trained the segmentation networks, leaving 429 patients (one patient had missing prostate contours) for the equivalence analysis. 50 more were excluded from the predictive analysis
because no pre-treatment IPSS was available, leaving 380 patients (Table~\ref{tab:cohort}). Missing covariates were imputed to the cohort median (per-fold) and an 80\,\% availability threshold was set for a variable to be used in the models.

The prediction target was $\Delta\mathrm{IPSS}=\mathrm{IPSS}(t)-\mathrm{IPSS}_{0}$, with $t=0$ the treatment day, at six
post-implant horizons : 30, 400, 730, 1125, 1460 and 1865 days post-treatment, each with its own
tolerance window. Within a window the measurement closest to the nominal horizon
was used, outside of it the value was estimated from a linear mixed model (Supplementary~\ref{sup:endpoints}). All metrics were computed on measured targets only, because
model-imputed targets are smooth by construction and inflate apparent
performance, leaving 354, 273, 199, 143, 187 and 209 patients for the respective endpoints evaluation.

\begin{table}[width=\linewidth,cols=4,pos=H]
  \centering
  \caption{Cohort characteristics of the predictive analysis set ($N=380$).
    Continuous variables are given as mean $\pm$ standard deviation, categorical
    variables as count (\%). Dose-volume indices are listed without summary
    statistics because their values depend on the segmentation source. Negative
    values for the time to first IPSS measurement denote pre-treatment
    measurements.}
  \label{tab:cohort}
  \begin{tabular}{@{}llrr@{}}
    \toprule
    Variable & Units & $n$ & Mean $\pm$ s.d. \\
    \midrule
    \multicolumn{4}{@{}l}{\textit{Continuous variables}} \\
    Age & years & 380 & $63.74 \pm 7.17$ \\
    ISUP grade &  & 323 & $1.18 \pm 0.39$ \\
    Pre-tx PSA & ng/mL & 380 & $5.62 \pm 2.99$ \\
    PSA density & ng/mL/cc & 350 & $0.16 \pm 0.10$ \\
    Post-tx prostate volume & cc & 350 & $35.89 \pm 10.49$ \\
    Pre-tx prostate volume & cc & 352 & $31.63 \pm 9.35$ \\
    Number of needles &  & 380 & $20.65 \pm 3.40$ \\
    Time until first IPSS measure & days & 380 & $-68.99 \pm 68.60$ \\
    Number of pre-tx IPSS measures &  & 380 & $1.14 \pm 0.38$ \\
    Obstructive IPSS pre-tx score &  & 380 & $3.16 \pm 3.78$ \\
    Irritative IPSS pre-tx score &  & 380 & $3.83 \pm 2.98$ \\
    \midrule
    \multicolumn{4}{@{}l}{\textit{Categorical variables}} \\
    Clinical T stage: T1C / T2A / T2B / T2C / T1B & & 380 & 74.2 / 22.4 / 2.9 / 0.3 / 0.3\,\% \\
    Gleason score: 6 / 7 / 8 & & 380 & 69.5 / 30.0 / 0.5\,\% \\
    Hormone therapy (ADT) & & 380 & 15.0\,\% \\
    DIL boost & & 359 & 8.2\,\% \\
    \midrule
    \multicolumn{4}{@{}l}{\textit{Dose-volume indices entering the models}} \\
    \multicolumn{4}{@{}l}{Prostate $D_{90}$, $V_{100}$, $V_{150}$, $V_{200}$} \\
    \multicolumn{4}{@{}l}{Urethra $D_{5}$, $D_{10}$, $D_{30}$, $D_{0.1\mathrm{cc}}$} \\
    \multicolumn{4}{@{}l}{Bladder neck $D_{1\mathrm{cc}}$, $D_{2\mathrm{cc}}$, $V_{100}$} \\
    \bottomrule
  \end{tabular}
\end{table}

\subsection{Segmentation sources and dose-volume indices}\label{subsec:seg_dvh}

Three contour sources were compared:  the reference expert manual delineation on the post-implant CT, a deterministic nnUNet, and a Bayesian nnUNet. The Bayesian conversion of the deterministic nnUNet was performed with the \texttt{torchbayesian} ~\cite{TorchbayesianTorchbayesian2026} library. More details on the Bayesian conversion are available in Supplementary~\ref{sup:bayesian}. At inference, $M=20$ passes were
drawn per case with the weights sampled from the trained posterior~\cite{sahlstenApplicationSimultaneousUncertainty2023}. Each DVH index was computed independently on every draw, the
expectation serving as the point estimate and the variance as an optional
predictive feature (Supplementary~\ref{sup:bayesian}). The dose distribution was fixed
across all three sources. No DVH index difference can arise from dose
recalculation.

Both networks were mainly trained on external images datasets: the LUND-PROBE~\cite{rogowskiLUNDPROBELUNDProstate2025}
and Prostate-Anatomical-Edge-Cases~\cite{thompson2023stress} collections. 18 post-implant
CT cases from this cohort were added to the training set, as the only images carrying radioactive seeds. The training distribution
therefore under-represents the post-implant setting, which biases the comparison
against the automatic sources (Supplementary~\ref{sup:bayesian}). 

Two structures are excluded from the equivalence analysis by construction: the
urethra, manually delineated in all three pipelines with a urinary catheter in place so that its indices are
identical across sources, and the bladder neck, built geometrically from the
prostatic base following Hathout et al.~\cite{hathoutDoseBladderNeck2014}, so its variation comes
from the prostate alone. The equivalence analysis bears on the prostatic indices of Table~\ref{tab:cohort} alone and the predictive analysis on the full eleven-DVH index panel. Doses were read according to Supplementary~\ref{sup:bayesian},
and expressed relative to the prescription. The panel of DVH indices was chosen from literature for their reported predictive power of urinary toxicities following LDR-BT treatment of prostate cancer~\cite{hathoutDoseBladderNeck2014,keyesPredictiveFactorsAcute2009a,tanimotoPredictiveFactorsAcute2013,moriPredictiveFactorsProlonged2017,djemhiUrethralDosimetryUrinary2026,farrisPatientAssessmentLower2021} (Supplementary~\ref{sup:bayesian}).

\subsection{Cohort-level equivalence}\label{subsec:methods_equiv}

To test for equivalence rather than the absence of difference, we used two one-sided tests on the paired differences
$\Delta_{i}=\mathrm{index}^{\mathrm{manual}}_{i}-\mathrm{index}^{\mathrm{auto}}_{i}$,
with non-equivalence as the null, $H_{0}:\lvert\mu\rvert\ge\delta$. Equivalence
was declared when the 90\,\% confidence interval of the mean difference ($\bar{d}$) fell
entirely within $[-\delta,+\delta]$. DVH indices from manual contours were compared to the ones from both deterministic and Bayesian (expectation taken as the point estimate) nnUNet automatic-segmentation networks.

DVH indices margins were fixed before any difference was examined and set below one standard
deviation of the inter-observer contouring noise floor. An automatic source is thus declared equivalent only
if it shifts an index by less than the dispersion between experts. Only the contouring arm of
published uncertainty and only CT-derived figures were kept, this cohort
being entirely post-implant CT~\cite{leeInterobserverVariabilityLeads2002,hanEffectInterobserverDifferences2003,debrabandereProstatePostimplantDosimetry2012,xueEffectInterobserverVariability2006,sandersUncertaintyMagneticResonance2023a}. The equivalence margins are 10\,\% of prescription for
$D_{90}$, 3\,\% of volume for $V_{100}$, 4\,\% for $V_{150}$ and
$V_{200}$. The $V_{200}$ margin is extrapolated from the $V_{150}$ value and our cohort's isodose ratio since no published figure is available
(Supplementary~\ref{sup:margins}). Cohort-level equivalence addresses specifically the mean difference $\bar{d}$ over the whole cohort. Individual agreement is described by the patient-level
equivalence rate,
$\mathrm{E}_{i}=\mathbf{1}\{\lvert\Delta_{i}\rvert\le\delta\}$, and by Lin's
concordance correlation coefficient~\cite{linConcordanceCorrelationCoefficient1989}.

\subsection{Patient-level quality thresholds}\label{subsec:methods_thresholds}

At the level of a single patient, the objective becomes to identify a segmentation quality threshold, above which a DVH index derived from an automatic contour can be considered equivalent ($E_i=1$). A logistic
regression of $\mathrm{E}$ on a segmentation quality score was fitted for each DVH index at
their respective margins and read at an equivalence probability $p^{*}=0.90$, with a
bootstrap interval over patients. We empirically established the segmentation quality metric as the Dice coefficient,
by comparing eight candidate metrics spanning overlap, distance and volumetric
families as predictors of $\mathrm{E}$ (Supplementary~\ref{sup:margins}).

The regression model was then modified to include the prostate's clinically recorded volume, in addition to the Dice coefficient, through an additive and an interactive (volume$\times$Dice) term. Only the
interaction was subject to selection, retained at $p<0.01$ on a likelihood ratio
test against the additive form because its coefficient enters the threshold expression's
denominator. 

\subsection{IPSS prediction models}\label{subsec:methods_models}

The baseline model $M_{0}$ contained all the clinical
variables of Table~\ref{tab:cohort}. Four feature blocks, containing the DVH panel derived
from the manual, deterministic and Bayesian sources ($\mathbb{E}[\mathrm{index}]$ only and $\mathbb{E}[\mathrm{index}]$ with across draw variance), were added to the clinical variables in $M_1$ models. Models were evaluated by nested cross-validation~\cite{varmaBiasErrorEstimation2006} with five outer ($K=5$) and four inner folds. Hyperparameters were
tuned with Optuna~\cite{akibaOptunaNextgenerationHyperparameter2019} within the inner loop of each outer training set. The outer
partition was repeated five times, giving 25 outer folds (Supplementary~\ref{sup:power}, Fig.~\ref{figS:nested_cv}). Partitions and seeds were identical across conditions so that
out-of-fold predictions are pairable patient by patient. ElasticNet~\cite{zouRegularizationVariableSelection2005} was specified as the principal learner. Random forest~\cite{biauRandomForestGuided2016}, XGBoost~\cite{chenXGBoostScalableTree2016}, CatBoost~\cite{prokhorenkovaCatBoostUnbiasedBoosting2018} and a
multilayer perceptron served as a robustness panel. A binary transposition used
the one-sided success label $\Delta\mathrm{IPSS}\le3$, the minimal clinically
important difference~\cite{windischConvertingInternationalProstate2024,babarPredictorsAchievingMinimal2024}, at every measurement between 1 and 5 years.

\subsection{Statistical analysis}\label{subsec:methods_stats}

Analyses used Python (3.12.3). $p$-values
are from two-sided tests at the 5\,\% level except the two one-sided equivalence
tests of the cohort-level equivalence analysis. Holm-Bonferroni correction~\cite{holmSimpleSequentiallyRejective1979} was applied to the cohort-level equivalence analysis. Normality assumptions were verified from distribution skewness, kurtosis and
quantile-quantile plots instead of normality pre-tests. Each conclusion
relying on a normal approximation was tested with a distribution-free
alternative (Supplementary~\ref{sup:distrib}).
 
The predictive analysis uses the difference in
root mean squared error ($\mathrm{RMSE}$) between $M_0$ and the different variations of $M_1$,
$\Delta\mathrm{RMSE}=\mathrm{RMSE}(M_{0})-\mathrm{RMSE}(M_{1})$, computed on the
out-of-fold predictions. Positive values of $\Delta\mathrm{RMSE}$ indicate better predictive performance. A percentile bootstrap 95\,\% confidence interval over
\num{5000} replicates is computed. It is complemented by a corrected
Nadeau-Bengio resampled $t$-test~\cite{nadeauInferenceGeneralizationError2003} on the per-fold differences, because the naive paired
$t$-test is anti-conservative since training sets overlap across folds~\cite{NIPS2003_e82c4b19} (Supplementary~\ref{sup:power}).
 
To declare significance, three criteria needed to be met: a
bootstrap interval excluding zero, a corrected $p$-value below 0.05, and a consistent sign of $\Delta\mathrm{RMSE}$ across the learner panel. The minimal clinically
important difference (MCID) of 3 IPSS points was converted into the RMSE gain
it would produce for $M_0$ at each horizon,
$\delta_{\mathrm{MCID}}=\mathrm{RMSE}(M_{0})-\sqrt{\mathrm{RMSE}(M_{0})^{2}-\mathrm{MCID}^{2}}$
(Supplementary~\ref{sup:headroom}). Positive controls ablate the
pre-treatment IPSS, full score or obstructive subscore alone, negative controls
ablate age and prostate volume. Analysis code is publicly available (\url{https://github.com/lb0918/Dosimetric_equivalence_prostate_automatic_contours}). 

\section{Results}\label{sec:Results}

\begin{figure}[pos=H]
    \centering
    \includegraphics[width=0.8\linewidth]{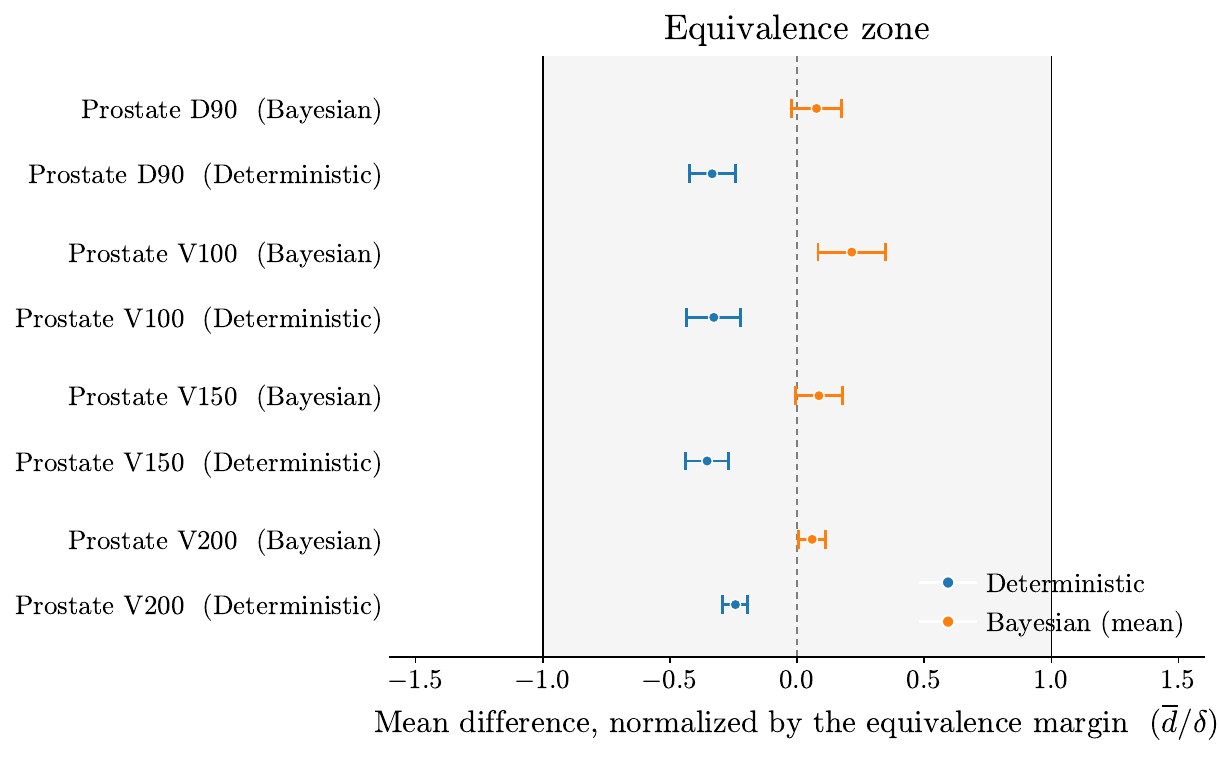}
    \caption{Cohort-level equivalence of the prostatic dose-volume indices. Each
    90\,\% confidence interval on the mean manual-minus-automatic difference is
    divided by its own pre-declared margin, so that the outer end of an interval
    gives the fraction of that margin actually consumed and equivalence
    corresponds to the interval lying within $\pm1$.}
    \label{fig:cohort_level_equ}
\end{figure}

All eight
comparisons, of the four DVH indices from the two automatic-segmentation sources, were declared equivalent at the cohort level, with $p_{\mathrm{TOST}}<10^{-19}$ throughout
(Table~\ref{tab:concordance}, Fig.~\ref{fig:cohort_level_equ}). The
deterministic network over-estimated the cohort's $D_{90}$ values by $3.34\%$ and by $0.97\%$ to $1.41\%$ on the three $V_{x}$ indices. All four 90\,\%
confidence intervals exclude zero. The Bayesian expectation reduces the cohort's mean difference error to between $+0.24\%$ and $+0.77\%$ across indices. No interval consumes more than 44\,\% of
its equivalence margin. The Bayesian expectation is nonetheless the more dispersed
estimator, with standard deviation of 12.53 against 11.29 (deterministic) percentage points of prescription on $D_{90}$,
and its concordance coefficient is consequently lower.

\begin{table}[width=\linewidth,cols=8,pos=H]
  \centering
  \caption{Equivalence of the automatic dose-volume indices with the expert
reference. Cohort-level columns give the mean bias with its 90\,\% confidence
interval and Lin's concordance correlation coefficient (CCC). Patient-level
columns give the proportion of patients whose individual difference satisfies
the equivalence margin $\delta$ and, for the deterministic source only, the
logistic Dice threshold at $p^{*}=0.90$ with its bootstrap interval and the
area under the curve (AUC) of the raw Dice as a threshold-free classifier of
equivalence. The $^*$ over the $V_{100}$ Dice threshold indicates that this value is above the highest Dice score observed in our cohort.}
  \label{tab:concordance}
  \small
  \setlength{\tabcolsep}{4pt}
    \begin{tabular}{lclccccc}
    \toprule
     & & & \multicolumn{2}{c}{Cohort level} & \multicolumn{3}{c}{Patient level} \\
    \cmidrule(lr){4-5}\cmidrule(lr){6-8}
    DVH index & $\delta$ & Source & Mean bias~[90\,\% CI] & CCC
      & Equiv. rate & Dice threshold & AUC \\
    \midrule
    $D_{90}$  & 10 & Bayesian      & $+0.77$\ \ $[-0.23,\,+1.76]$ & 0.627 & 66.1\,\% & -- & -- \\
              &    & Deterministic & $-3.34$\ \ $[-4.24,\,-2.44]$ & 0.661 & 68.6\,\% & 0.925 $[0.910,\,0.945]$ & 0.782 \\
    \addlinespace
    $V_{100}$ & 3  & Bayesian      & $+0.65$\ \ $[+0.25,\,+1.04]$ & 0.759 & 55.9\,\% & -- & -- \\
              &    & Deterministic & $-0.98$\ \ $[-1.30,\,-0.66]$ & 0.826 & 62.9\,\% & 0.958$^{*}$ $[0.933,\,0.995]$ & 0.743 \\
    \addlinespace
    $V_{150}$ & 4  & Bayesian      & $+0.34$\ \ $[-0.03,\,+0.71]$ & 0.928 & 71.1\,\% & -- & -- \\
              &    & Deterministic & $-1.41$\ \ $[-1.75,\,-1.08]$ & 0.934 & 72.2\,\% & 0.922 $[0.906,\,0.945]$ & 0.772 \\
    \addlinespace
    $V_{200}$ & 4  & Bayesian      & $+0.24$\ \ $[+0.03,\,+0.45]$ & 0.943 & 89.7\,\% & -- & -- \\
              &    & Deterministic & $-0.97$\ \ $[-1.17,\,-0.77]$ & 0.944 & 87.6\,\% & 0.864 $[0.848,\,0.880]$ & 0.789 \\
    \bottomrule
  \end{tabular}
\end{table}

Applied patient by patient, the equivalence margins are satisfied by 62.9 to 87.6\,\%
of patients depending on the DVH index, for the deterministic source
(Table~\ref{tab:concordance}). No competing segmentation quality metric outperformed Dice for
discrimination (Supplementary~\ref{sup:margins}).
 
\begin{figure}[pos=H]
    \centering
    \includegraphics[width=0.72\linewidth]{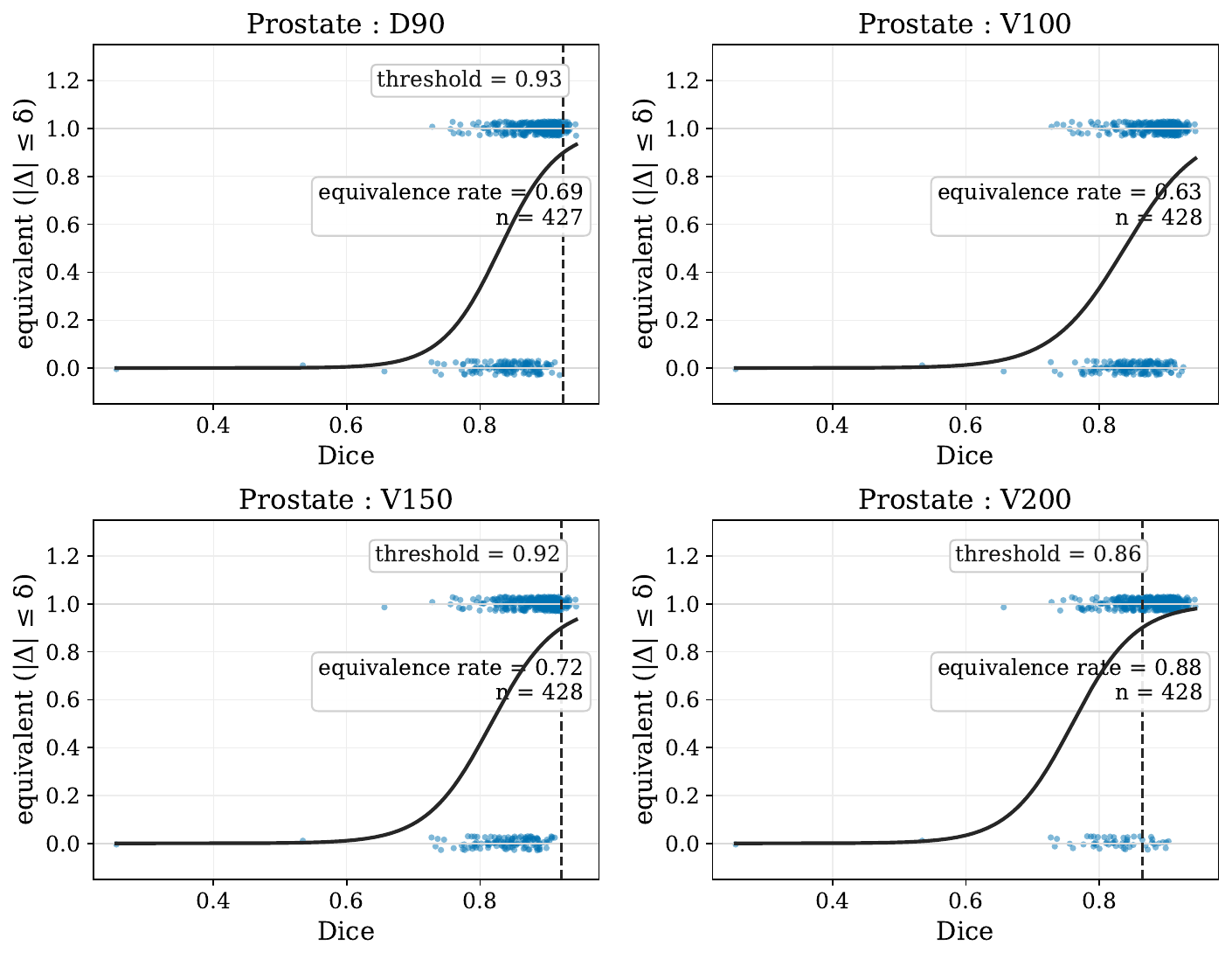}
    \caption{Patient-level equivalence threshold on the Dice coefficient at the margins specified in Table~\ref{tab:concordance}. The classifier is a logistic regression, which gives a probability of equivalence, with the
    threshold at $p^{*}=0.90$. The $V_{100}$ panel carries no marker because its
    threshold falls above the observed Dice range.}
    \label{fig:dice_threshold}
\end{figure}

Read at $p^{*}=0.90$, the volume-independent estimator returns thresholds between 0.864 and 0.958
(Table~\ref{tab:concordance},
Fig.~\ref{fig:dice_threshold}). The $V_{100}$ threshold exceeds the
highest Dice this pipeline attains, 0.944, and is flagged as
extrapolated. This means that a 90\,\% probability of equivalence for $V_{100}$ from automatic-deterministic contours would apply to no patient in this cohort, based on Dice scores.

\begin{figure}[pos=H]
    \centering
    \includegraphics[width=0.73\linewidth]{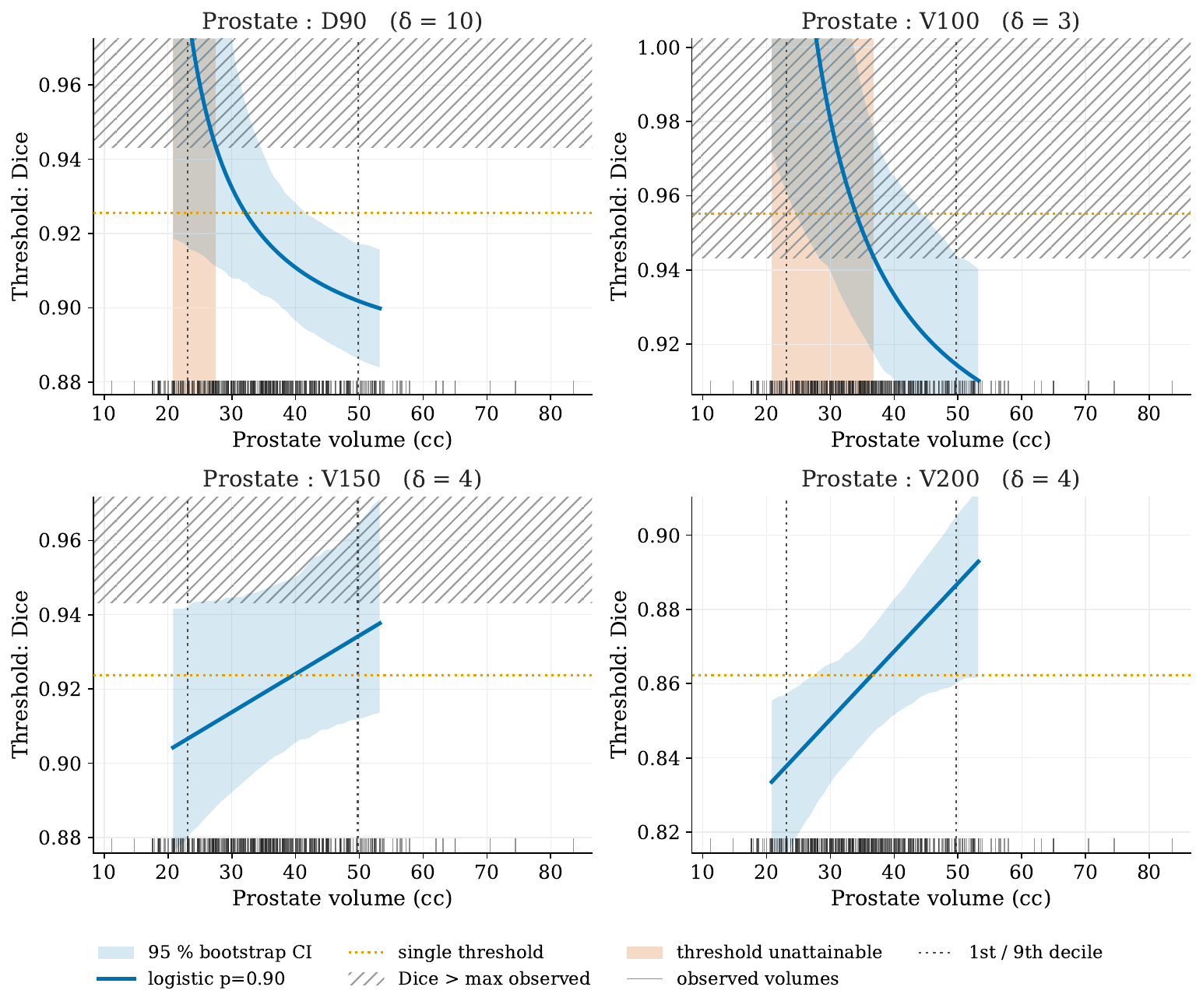}
    \caption{Dice threshold for patient-level equivalence as a function of
    prostate volume. The solid curve is a logistic regression at $p^{*}=0.90$
    with its 95\,\% bootstrap interval, drawn between the 5th and 95th percentiles
    of the observed volumes in this cohort. The dotted horizontal line is the
    volume-independent threshold. Rug marks show the prostate volumes
    observed in the cohort, and the dotted vertical lines their first and ninth
    deciles. The hatched area lies above the highest Dice observed in the cohort
    (0.944). The shaded volume interval indicates where the threshold falls in
    this area, making it unattainable with our deterministic segmentation
    network.}
    \label{fig:dice_vol_threshold}
\end{figure}

In our cohort, Dice is correlated with
prostate volume ($\rho=0.311$), because a fixed contouring
error costs more overlap on a small gland~\cite{johnssonAnalyticalPerformanceAPROMISE2022}. For the volume-dependent estimators, the interaction was retained on
$D_{90}$ and $V_{100}$ and the additive form on $V_{150}$ and $V_{200}$. Where the interaction term is retained the threshold falls steeply as the
gland grows (Fig.~\ref{fig:dice_vol_threshold}). For $D_{90}$ the threshold goes from $0.980$ at the
first decile of volume (23.1~cc) to $0.902$ at the ninth (49.8~cc). Below roughly
27~cc, no attainable Dice reaches the equivalence probability $p^{*}$ for
$D_{90}$ (shaded interval in Fig.~\ref{fig:dice_vol_threshold}). On
$V_{100}$, $p^{*}$ stays unattainable until roughly 37~cc. The two additive curves rise
instead, from 0.907 to 0.934 on $V_{150}$ and from 0.838 to 0.887 on $V_{200}$,
so the sign of the volume effect is not shared by the four indices.

The clinical baseline explains an increasing share of the variance of
$\Delta\mathrm{IPSS}$ along the horizons, $R^{2}$ rising from 0.142 at 1 month
to 0.390 at 3 years and its $\mathrm{RMSE}$ falling from 6.74 to 4.97
IPSS points. At every horizon the $\mathrm{RMSE}$ sits 1.17 to 4.43 points above the measurement floor of the survey, so a predictor carrying meaningful information would
have provided predictive improvement (Supplementary~\ref{sup:headroom}).
 
\begin{figure}[pos=H]
    \centering
    \includegraphics[width=\linewidth]{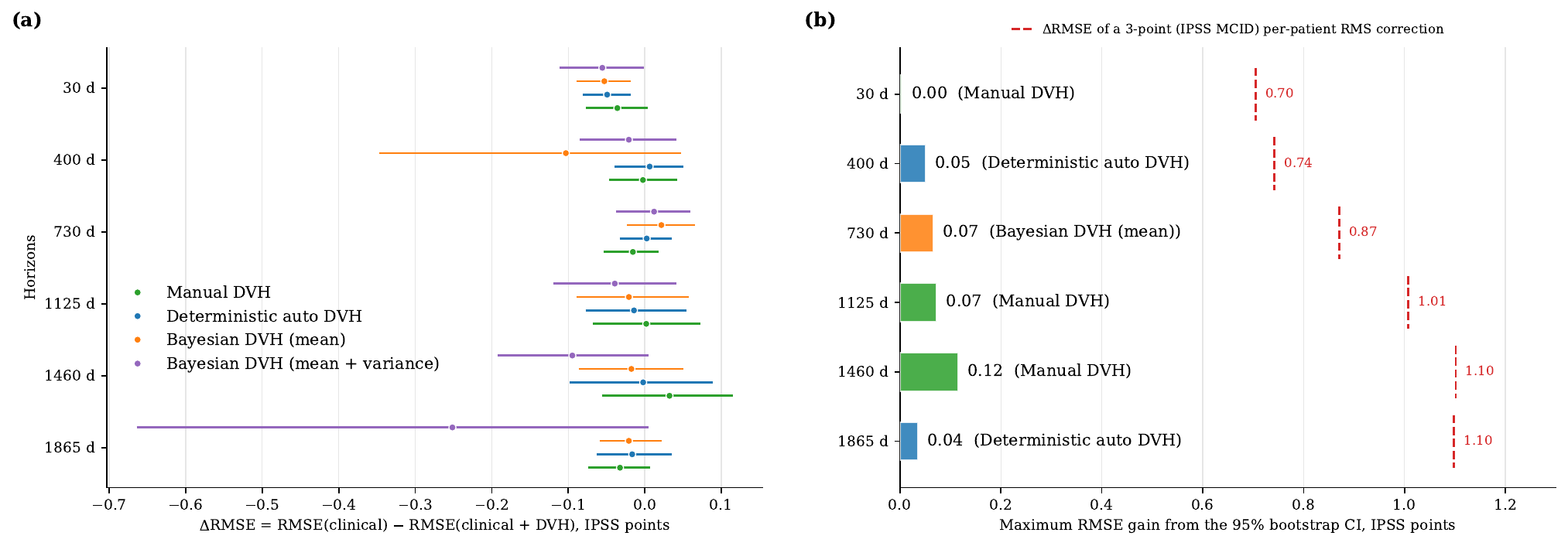}
    \caption{Incremental value of the dose-volume block over the clinical
    baseline for the prediction of $\Delta\mathrm{IPSS}$ with
    ElasticNet learner. (a) $\Delta\mathrm{RMSE}$ with its 95\,\%
    bootstrap confidence interval, for the four feature blocks at the six
    horizons. Positive values indicate improvement. (b) Most favourable upper
    confidence limit over the four blocks at each horizon, giving the largest
    improvement the data remain compatible with. The red dashed marks give, at each horizon, the RMSE gain that a correction of
the baseline predictions with root-mean-square amplitude equal to the MCID
(3 IPSS points) would produce.}
    \label{fig:IPSS_continuous_pred_IC}
\end{figure}

Across the twenty-four tests, four DVH blocks (one per segmentation source + Bayesian variance) at six horizons with the ElasticNet, no cell satisfied the declared criteria for significance
(Fig.~\ref{fig:IPSS_continuous_pred_IC}a). Three bootstrap intervals excluded
zero, all at 1 month and all in the direction of degradation, the
widest being $-0.055$ $[-0.112,\,-0.001]$. None was confirmed by the corrected test, whose smallest
$p$-value was 0.120, and the median absolute difference over the
twenty-four cells is 0.02 IPSS points. Taking at each horizon
the most favourable upper limit over the four blocks of the ElasticNet model
(Fig.~\ref{fig:IPSS_continuous_pred_IC}b), the largest improvement is 0.12 IPSS points at 4 years, against 1.10 for a correction
equal to the MCID at the same horizon.
 
Over five algorithms, the DVH blocks yield at most two bootstrap intervals
excluding zero in the way of improvement ($\Delta\mathrm{RMSE}>0$) for any given block and horizon (all at 5 years with the RF and XGBoost models), and never with a corrected
$p$-value below 0.05. Furthermore, no DVH block shifts the AUC of the binary target by more than $+0.02$. Ablating the pre-treatment total
IPSS costs $+0.38$ to $+1.46$ IPSS points and the obstructive subscore alone
$+0.23$ to $+0.79$, both declared significant at nearly every horizon, while the negative
controls are never declared significant. For the binary target, total pre-treatment IPSS is declared significant but the obstructive
subscore is declared no more often than age or prostate volume (Supplementary~\ref{sup:panel}).

\section{Discussion}\label{sec:discussion}
The equivalence analysis suggests a reproducible procedure rather than declaring two particular segmentation networks as dosimetrically equivalent to a human expert. Its three parts are each editable without altering the logic.
The equivalence margin $\delta$ represents the dosimetric shift already tolerated between experts, derived from human contouring IOV. The segmentation quality metric was established empirically in our cohort, but could be replaced by other metrics. Finally, the probability $p^*$ at which the patient-level
threshold is read can be changed freely. Nothing in that structure is specific to LDR-BT dosimetry or to the CT imaging modality. A centre using another
segmentation network, tolerance margins or segmentation quality metric can thus apply a similar procedure and obtain an equivalence
criterion in its own terms~\cite{vanaalstDosebasedEvaluationDelineation2026,heilemannClinicalImplementationEvaluation2023}. Our work also supports clinical evaluation of automatic contours which requires the valid assessment of their dosimetric consequences alongside geometric correspondence. While describing complementary aspects of contour agreement, typical metrics, such as Dice and HD95, do not establish the equivalence of resulting DVH indices when compared to expert-generated delineations. From a clinical perspective, accurate characterisation of the dose distribution within the target and surrounding organs at risk, commonly summarised by DVH indices, is essential. Its valid evaluation requires an appropriate statistical framework; the failure to detect a significant difference does not establish equivalence. Equivalence testing instead assesses if the confidence interval for the mean dosimetric difference lies entirely within appropriately predefined margins. Anchoring these margins to human IOV provides an explicit reference for evaluating automatic contours.

The equivalence margins we chose are set below the lowest published standard deviation of IOV, for the CT modality in an effort to be conservative in the equivalence verdict. Even then, the confidence intervals of the cohort-level analysis never consumed more than 44\% of the equivalence margins, which indicates that, in our cohort, margins twice as strict would have allowed for a cohort-level equivalence verdict for all eight comparisons. Our margins are based on published multi-institutional panels, which can inflate the IOV. Each centre should determine its own IOV on specific DVH indices by comparing the values of segmentations from their own clinicians. 

Nonetheless, a cohort verdict bears on a mean, so equivalence on average carries no implication for
the patient in front of the physician. A patient-level threshold is necessary to tell what
the cohort-level analysis cannot. On $D_{90}$ and $V_{100}$, a
fixed contouring error costs more overlap on a small gland, so the quality
required for patient-level equivalence rises as the prostate shrinks, whereas on the two
high-dose volumes the dependence is additive and increases with prostate volume. The
direction of that dependence is index-specific and has to be empirically estimated rather
than assumed~\cite{al-qaisiehImpactProstateVolume2002}. A quality score is not available prospectively, since computing it requires the expert contour the automatic one is meant to replace. The patient-level
thresholds are therefore limited to a commissioning tool and not a per-patient acceptance
test. Taking the Dice-volume distribution of a validation cohort
$f(D\mid v)$ and combining it with the logistic model gives the expected rate of equivalence at each
prostate volume, $P(E\mid v)=\int P(E\mid D,v)\,f(D\mid v)\,\mathrm{d}D$. Since
prostate volume is known before post-implant dosimetry, this profile can flag
in advance the patients whose automatic contours are unlikely to be equivalent,
but it cannot detect dosimetric equivalence of a given contour. True patient-level equivalence would require a
reference-free quantity calibrated against equivalence. The across-draw
dispersion of the Bayesian network discriminated poorly in our data
(Supplementary~\ref{sup:margins}), and its calibration is an out-of-distribution (OOD) detection problem that
warrants a separate study.

For the predictive analysis, the DVH indices add negligible predictive power to a clinical baseline
for $\Delta\mathrm{IPSS}$ in our cohort at the specified horizons, irrespective of the segmentation source. At every horizon the upper confidence
limit excludes a $\Delta\mathrm{RMSE}$ that reaches $\delta_{\mathrm{MCID}}$, which makes this null result bounded and robust. The corrected $t$-tests and the learner panel behaviour both also indicate that the DVH block does not improve prediction. It remains a statement about this
cohort, these endpoints and these eleven scalars, and does not guarantee a general
claim that implant dosimetry is uninformative about urinary toxicity. Treatment planning enforces tolerance values on the DVH indices, so the delivered plans occupy a narrow band of the achievable dosimetric range. The null therefore applies to the dosimetric spread that our cohort's clinical planning actually produces rather than to dose-volume effects in general. Cohorts planned under comparable constraints have nonetheless reported significant predictive power~\cite{hathoutDoseBladderNeck2014,keyesPredictiveFactorsAcute2009a,djemhiUrethralDosimetryUrinary2026}. Since planning is performed on the day of implantation whereas dosimetry is evaluated a month later, prostate edema and source migration can separate the delivered indices from the planned ones and widen the range over which a signal could have been detected. Post-implant CT is a single snapshot of a dose distribution
delivered over several months. About 30\,\% of the $^{125}$I dose is delivered
before the first month, when the prostate is still affected by edema, so the indices describe the
early period less accurately. Edema magnitude
and resolution vary widely between
patients~\cite{nathAAPMRecommendationsDose2009}, inducing an
error on the indices and harming predictions. Our conclusions rely solely on day-30 indices, as used clinically, not on the biologically
delivered dose. The four urethral indices are further removed from the delivered dose, being read on a urethra displaced and deformed by the catheter present only for imaging. The reliance on CT-based delineation is a limitation of this historical cohort. The limited soft-tissue contrast of CT compared with MRI may introduce uncertainty into the manual reference contours and the resulting DVH indices. This uncertainty may affect the assessment of dosimetric equivalence and attenuate the incremental predictive value of DVH indices for $\Delta\mathrm{IPSS}$. The temporal span of our study (2004-2024) also allows for heterogeneity in CT quality and delineation practices, potentially leading to missed dose-response relationships. However, we believe that the availability of longitudinal clinical and dosimetric data provides an invaluable opportunity to evaluate the proposed framework within an established CT-based workflow, which could subsequently be extended to MRI-based workflows, with appropriate validation and reassessment of the equivalence margins. The current work also remains relevant to centres where post-implant MRI is not routinely available.
 
The null predictive result nonetheless carries a methodological warning. Paired
$t$-tests using the outputs of predictive models trained with cross-validation violate the independence of observations, since any two outer training sets share patients. The naive
standard error therefore understates the variance of the mean per-fold
difference~\cite{NIPS2003_e82c4b19,nadeauInferenceGeneralizationError2003}, which could lead to false positives. Using a single family of predictive model is also imprudent since, across the
learner panel, individual cells do return bootstrap intervals excluding zero, but without sign consistency across learners and without survival of
the corrected test (Supplementary~\ref{sup:panel}). Had only one learner been fixed in advance and
the correction omitted, several of those cells would have been declared
significant. What prevents this is the conjunction of criteria, and the same
conjunction should be met before any variable is admitted as predictive~\cite{cabitzaWhyAlmostAll2026,collinsTRIPOD+AIStatementUpdated2024}.

The predictive analysis focused on IPSS because it provided longitudinal patient-reported outcomes (PROs) in this historical cohort. Comparable GI PRO data were lacking, and clinician-reported rectal toxicity records were insufficiently complete, as well as generally infrequent, for reliable analysis. More systematic EPIC-26 collection will support future evaluation of rectal dose/outcomes relationships and additional PRO domains.

\section{Conclusion}\label{sec:Conclusion}
This work provides a framework for declaring dosimetric equivalence of DVH indices, computed from automatic-segmentation with the ones computed on expert manual delineation. This method is entirely transferable to a different institution with different standards, imaging modality, or automatic segmentation algorithm. It is the dose within contours which matters in radiation oncology, not the geometrical agreement between the contours themselves, equivalence rather than difference should be studied. The proposed procedure, as presented here, could provide a cohort-level commissioning criterion, by combining a trained patient-level logistic model with a known Dice-volume distribution. Although a patient-level prospective use is not possible in the current format, further work directly linking the Bayesian dosimetric uncertainty to the equivalence probability could bridge this gap.

This study also explores the incremental predictive value of DVH indices, to a clinical baseline for $\Delta\mathrm{IPSS}$, depending on the segmentation sources, at six horizons, spanning a period of five years after implants. Even though the DVH block would have been declared significant at one horizon with a single prediction model and without Nadeau-Bengio's correlation correction, this behaviour could not be reproduced across models or hold against corrections. Adequate corrections to Student's test as well as multiple significance criteria and a variety of models should be used in such machine learning studies to safeguard against false positive results. All conclusions from the predictive analysis are limited to this cohort, these endpoints and the chosen DVH panel. It should not be interpreted as a claim that DVH indices and urinary toxicities in general are not related to one another.

\printcredits
\section*{Funding}
No funding was received for this project.\\
\textbf{Data availability}
The datasets generated and/or analyzed during
the current study are not publicly available due to patient privacy
considerations and institutional ethics restrictions. De-identified data
may be made available from the corresponding author upon reasonable
request, subject to approval by the relevant institutional research ethics
boards and data-sharing agreements where applicable.
\section*{Declaration of competing interests}
No author has any conflict of interest to declare, regarding this work.
\section*{Acknowledgements}
We want to thank all the students, physicists and physicians who participated in the collection and curation of the data used in this work. We specifically thank Raphaël Brodeur for useful discussions regarding Bayesian neural networks and Meriem Hassaine for data collection. We also want to thank the Department of Hematology, Oncology, and Radiation Physics at the Sk\aa{}ne University Hospital for sharing their synthetic CT dataset, on which our automatic segmentation models were trained.

\section*{Declaration of generative AI and AI-assisted technologies}

During the preparation of this work the authors used Claude (Anthropic)
to assist with the development of the analysis code and core text length reduction. The first author
reviewed and validated all generated code and text and takes full responsibility
for the content of the published article.

\clearpage
\etocdepthtag.toc{supplement}
\setcounter{section}{0}
\setcounter{figure}{0}
\setcounter{table}{0}
\setcounter{equation}{0}
\renewcommand{\thesection}{S\arabic{section}}
\renewcommand{\thefigure}{S\arabic{figure}}
\renewcommand{\thetable}{S\arabic{table}}
\renewcommand{\theequation}{S\arabic{equation}}
\renewcommand{\theHsection}{S\arabic{section}}
\renewcommand{\theHfigure}{S\arabic{figure}}
\renewcommand{\theHtable}{S\arabic{table}}
\renewcommand{\theHequation}{S\arabic{equation}}
\pdfbookmark[0]{Supplementary material}{supplementary}
\begin{center}
  {\sffamily\LARGE\bfseries Supplementary material}\par\medskip
  {\sffamily\large Dosimetric equivalence of prostatic DVH indices from automatic contours,\\
  with incremental predictive value for IPSS after LDR brachytherapy}\par
\end{center}
\bigskip
\etocsettagdepth{mainmatter}{none}
\etocsettagdepth{supplement}{subsection}
\etocsettocstyle{\noindent{\sffamily\bfseries Contents}\par\smallskip}{\par}
\tableofcontents
\clearpage

\section{Endpoint definition and measurement coverage}\label{sup:endpoints}

\begin{figure}[pos=H]
  \centering
  \includegraphics[width=0.9\linewidth]{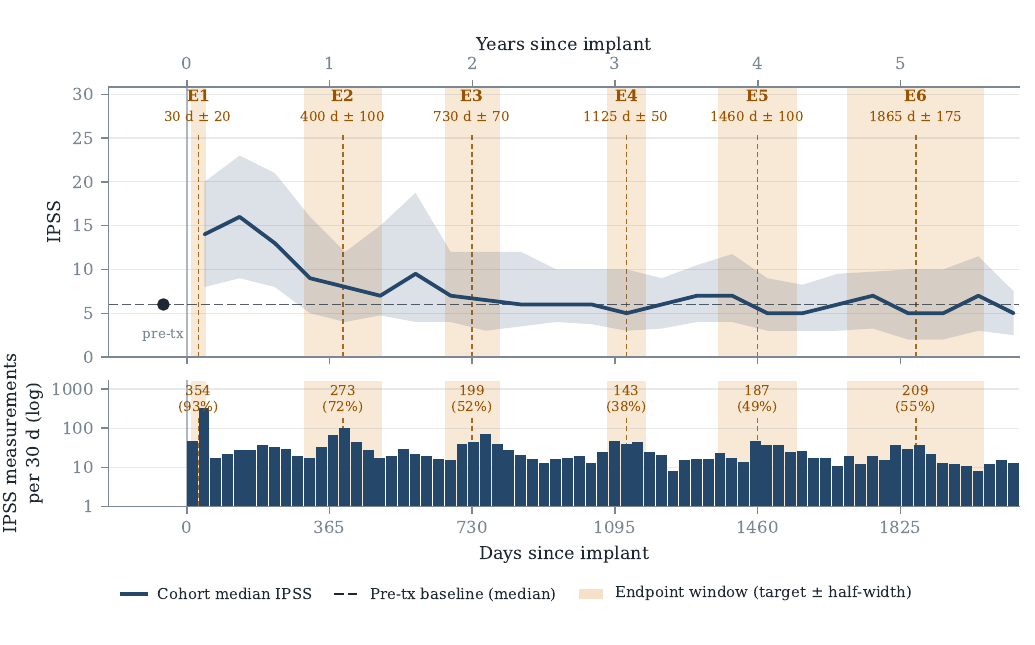}
  \caption{Horizon windows used in this work. Each horizon is shown with
    its tolerance half-window. The upper plot shows the evolution of the cohort IPSS scores. The lower plot shows the $\log$ of IPSS measurement frequency in time. The numbers in yellow represent the absolute number of IPSS measurements within a given window.}
  \label{figS:endpoint_windows}
\end{figure}

The six horizons are aligned with moments presenting the highest IPSS measurement densities in our cohort, as shown in Figure~\ref{figS:endpoint_windows}. Such a long coverage of IPSS evolution is not common in published studies~\cite{keyesPredictiveFactorsAcute2009a,tanimotoPredictiveFactorsAcute2013,moriPredictiveFactorsProlonged2017,farrisPatientAssessmentLower2021}.

When a measurement fell inside the tolerance window: 20, 100, 70, 50, 100 and 175 days for each horizon respectively, the measurement closest to
the center was used. Otherwise the endpoint value was estimated from a linear mixed model with a population-level
B-spline in time~\cite{perperoglouReviewSplineFunction2019} and a patient-level random intercept,
\begin{equation}\label{eqS:extrapolation_model}
  \mathrm{IPSS}(t)\sim \mathrm{bs}(t,\mathrm{df}=4)+(1\vert\mathrm{patient}).
\end{equation}
$t$ is the time since implant and the spline knots are placed at quantiles of
the measurement times. Estimated values were computed when the
horizon fell between a patient's first and last measurement. Outside that range the target was set to
missing.

Performance metrics were computed on real
targets only, because model-imputed targets can inflate model performance. The number of real targets was 354, 273, 199, 143, 187 and 209 respectively for the six horizons.

\section{Segmentation networks and dose-volume index computation}\label{sup:bayesian}
 
\subsection{Training data}
 
Both deterministic and Bayesian nnUNet were trained on a collection of two external datasets and a small dataset from our institution (CHU de Québec-Université Laval). Images from our institution used for segmentation model training were excluded from all subsequent analysis.
 
The first source is LUND-PROBE~\cite{rogowskiLUNDPROBELUNDProstate2025}, which has
432 patients from Sk\aa{}ne University Hospital, Lund. Each patient has a
T2-weighted MRI, a synthetic CT generated from it
and expert segmentations for the prostate, rectum, bladder, femoral
heads, penile bulb and genitalia. We only used the synthetic CTs to train our segmentation models, as our cohort's post-implant imaging modality is entirely CT.
The dataset is distributed under a data usage agreement through the AIDA Data
Hub.
 
The second is Prostate-Anatomical-Edge-Cases~\cite{thompson2023stress}, which has 131 patients with prostate, rectum and bladder delineation on CT. The dataset was
assembled to stress-test pelvic auto-segmentation: 112 of the 131 patients presented anatomical variants of different nature. The remaining 19 are a normal
comparison group. 

The third dataset is 18 post-implant CT cases from our institution, the
only training images that contain radioactive seeds and edema. The 18 corresponding patients were removed from the analysis cohort, as
stated in the main text.

Preprocessing, network topology and training schedule were left at the nnUNet~\cite{isenseeNnUNetSelfconfiguringMethod2021a}
defaults for the three-dimensional full-resolution configuration (\texttt{3d\_fullres}). The deterministic and Bayesian networks reached a mean Dice score of 0.863 and 0.859, respectively, on our test cohort (all our cohort minus the 18 used for segmentation training).

Two of the three sources of our training dataset are pre-implant, thus without seeds. LUND-PROBE's prostate contours are delineated on MRI, whereas CT-defined prostate
volumes are on average 32\,\% larger~\cite{roachProstateVolumesDefined1996}.
The training dataset therefore under-represents the images of our cohort. Domain gaps usually degrade the automatic contours relative to
what an in-domain training set would produce, and a worse automatic contour
should widen $\lvert\Delta\rvert$'s confidence interval. This domain gap therefore makes the equivalence verdicts of
the main text harder to obtain rather than easier, and the results are in that
sense more conservative.

\begin{center}
\begin{tabular}{@{}lrlllll@{}}
  \toprule
  Dataset & $n$ & Images & Seeds  & Use \\
  \midrule
  LUND-PROBE & 432 & synthetic CT & no\textsuperscript{a}   & training \\
  Prostate-Anatomical-Edge-Cases & 131 & planning CT & no   & training \\
  CHU de Québec & 18 & post-implant CT & yes   & training \\
  CHU de Québec & 429 & post-implant CT & yes   & evaluation \\
  \bottomrule
  \multicolumn{7}{@{}l}{\footnotesize\textsuperscript{a}Three gold fiducials per
  patient, rendered as artefact-free high-density spheres of 2.5~mm radius.}
\end{tabular}
\end{center}

\subsection{Variational conversion and fine-tuning}

We converted the trained deterministic nnUNet~\cite{isenseeNnUNetSelfconfiguringMethod2021a} into a Bayesian neural network by
variational inference, following Bayes-by-Backprop~\cite{blundellWeightUncertaintyNeural2015}. Each weight $w$ is replaced
by a Gaussian variational posterior $q(w)=\mathcal{N}(\mu,\sigma^{2})$, with
$\sigma=\log(1+e^{\rho})$, where $\mu$ and $\rho$ are trained parameters. The
conversion was performed with the \texttt{torchbayesian} library~\cite{TorchbayesianTorchbayesian2026}, which operates
at the level of individual \texttt{nn.Parameter} objects rather than by
substituting layer classes, so the nnUNet architecture itself is left unmodified.
All parameters of the network were made variational.

Training a Bayesian network from random initialisation is unstable and expensive,
so the Bayesian network was initialised from the converged weights of the
deterministic model
by the MOPED scheme~\cite{krishnanSpecifyingWeightPriors2020},
\begin{equation}\label{eqS:MOPED_scheme}
  \mu_{w}\leftarrow w_{\rm det},\quad
  \sigma_{w}\leftarrow\delta\lvert w_{\rm det}\rvert,\quad
  \rho_{w}\leftarrow\log(e^{\sigma_{w}}-1),
\end{equation}
with $\delta=0.5$. Variational training, from such an initialization, widens the
posterior distribution. Fine-tuning minimised the negative evidence lower bound,
\begin{equation}\label{eqS:loss}
  \mathcal{L}=\underbrace{\mathcal{L}_{\rm nnUNet}}_{\text{negative log-likelihood}}
  +\lambda_{\rm KL}\frac{\mathrm{KL}\!\left(q(w)\Vert p(w)\right)}{n_{\rm batches}},
\end{equation}
where the likelihood term is the unmodified nnUNet Dice plus cross-entropy loss
with deep supervision. The Kullback-Leibler term was normalised per batch and
annealed linearly from zero over the first 25 epochs. The network was fine-tuned for 250 epochs at a learning rate of
$1\times10^{-4}$.

\subsection{Inference protocol}

For each patient, $M=20$ independent inferences of the contours were produced, by sampling a different weight out of the posterior distribution. We
emphasise that this is not Monte Carlo dropout~\cite{sahlstenApplicationSimultaneousUncertainty2023}, since the stochasticity comes from the
variational posterior distribution.

Both trained networks came from the 
best validation iteration of the model, with the same nnUNet plan and configuration.
Deterministic prediction used the standard predictor with test-time augmentation
by mirroring, whereas Bayesian prediction was run with mirroring disabled. Augmentation-based spread is a heuristic
that measures sensitivity to a chosen set of geometric transformations, whereas
posterior sampling measures uncertainty about the weights themselves. Combining
the two would confound them within a single dispersion estimate.

\subsection{Dose-volume index computation}

All DVH indices were computed on a fixed dose distribution, regardless of contour source, so that any difference between the indices originates from the variability of the contours themselves.
Doses were read on their native grid and never resampled in-plane, since resampling creates smoothing near the sources where the gradients are steep in LDR-BT. 

Doses were clipped at 655.35~Gy in any given voxel. The cohort contains two DICOM dose encodings, one that saturates the
16-bit storage at that value and one that does not, reaching dose values orders of magnitude higher. This
is an artefact of the TG-43 dose computation formalism~\cite{kirisitsReviewClinicalBrachytherapy2014}. Left uncorrected, every
high-dose index would depend on the encoding rather than on the actual treatment. For the same reason, maximum-dose metrics ($D_{\max}$) were excluded
from the DVH panel in the predictive analysis. Percentile and threshold metrics below $D_{\max}$ are less affected
by this choice.

Indices follow standard conventions, with dose-at-absolute-volume indices
($D_{x\mathrm{cc}}$) expressed as a fraction of the prescribed dose. The DVH panel of the predictive analysis includes indices from three structures. Four
prostatic indices were chosen: $D_{90}$, $V_{100}$, $V_{150}$ and $V_{200}$.
For the urethra, $D_{5}$, $D_{10}$, $D_{30}$ and
$D_{0.1\mathrm{cc}}$ were retained. Since the bladder neck is approximately 2.2~cc~\cite{hathoutDoseBladderNeck2014},
$D_{2\mathrm{cc}}$ probes very nearly the whole structure for some patients. We thus decided to add $D_{1\mathrm{cc}}$ and $V_{100}$ as complementary indices to the DVH panel. All chosen indices, except for $D_{1\mathrm{cc}}$ and $V_{100}$ of the bladder neck, have been reported at least once as significant predictors of urinary toxicities following radiotherapy~\cite{keyesPredictiveFactorsAcute2009a,tanimotoPredictiveFactorsAcute2013,hathoutDoseBladderNeck2014,moriPredictiveFactorsProlonged2017,djemhiUrethralDosimetryUrinary2026, crookUtilityTransitionZone2002, cendalesOriginalPaperHighdoserateBrachytherapy2015,vecciaImpactPostimplantDosimetric2015}. 

The prescription dose was 144 or 145~Gy for every patient in this cohort. DVH indices were normalised by the prescription or the structure volume, since the published
inter-observer data on which the equivalence margins are based are
expressed as relative indices.

For the Bayesian source, indices were computed independently for each of the
$M=20$ draws. It gave us a distribution per patient/structure, which allowed us to compute the expectation as a point estimate value and the variance as a measure of uncertainty.

\section{Equivalence margins and threshold methodology}\label{sup:margins}

Equivalence margins were determined from the literature before any contours were examined, and their references are shown in Table~\ref{tabS:margins}. They represent the disagreement in DVH indices value that two human experts already produce between themselves~\cite{kirisitsReviewClinicalBrachytherapy2014,debrabandereProstatePostimplantDosimetry2012}. If replacing
expert manual delineation by automatic segmentation shifts an index by less than such a margin,
that shift cannot be the source of a different clinical decision, since the
decision is already indifferent to an error of that amplitude. Each margin was therefore set below one standard deviation of the
inter-observer contouring noise for the CT modality. 

Two considerations restrict which published figures can be considered as equivalence margins. Since the dose distribution is constant across contour sources, only the contouring arm of post-implant segmentation uncertainty can be considered. The
seed-reconstruction and fusion arms, which several sources report separately~\cite{debrabandereProstatePostimplantDosimetry2012,westendorpDosimetricImpactContouring2017,lindsaySystematicStudyImaging2003}, are
null by construction in our design and would inflate the margins if included. Because human contouring error follows the ordering TRUS $<$ MRI $<$ CT~\cite{westendorpDosimetricImpactContouring2017,lavoie-gagnonAdvantagesTRUSbasedDelineation2022,sandersComputeraidedSegmentationMRI2022,crookInterobserverVariationPostimplant2002b}, and this
cohort is entirely post-implant CT, we only considered CT values for margins. The figures for other modalities are shown only as comparisons. A margin calibrated on the MRI
or ultrasound IOV would impose on CT
contouring an unachievable reproducibility. Anchors reported as relative dispersions were converted to
the scale of our study using the median index value of our
cohort.

\begin{table}[pos=H]
  \centering
  \caption{Published inter-observer contouring dispersions and the pre-declared
    margins. Source figures, reported as relative dispersions, are converted onto
    this cohort's median index values (107.64, 93.08, 57.60 and 27.33 for $D_{90}$, $V_{100}$, $V_{150}$ and $V_{200}$ respectively).
    Dose indices are expressed in percentage points of the prescription dose,
    volume indices in percentage points of volume. A dash indicates that the source does not report the index, a
    $^{\dagger}$ marks a value extrapolated from the same source's $V_{150}$
    figure by the isodose gradient factor.}
  \label{tabS:margins}
  \small
  \begin{tabular}{llcccc}
    \toprule
    Source & Modality & $D_{90}$ & $V_{100}$ & $V_{150}$ & $V_{200}$ \\
    \midrule
    Sanders 2023, radiation oncologists~\cite{sandersUncertaintyMagneticResonance2023a} & MRI  & 4.9  & 2.2  & 1.7 & 1.0$^{\dagger}$ \\
    Xue 2006~\cite{xueEffectInterobserverVariability2006}                              & TRUS & 8.5  & 3.7  & --  & -- \\
    Han 2003~\cite{hanEffectInterobserverDifferences2003}                              & CT   & 10.9 & 3.7  & --  & -- \\
    Lee 2002~\cite{leeInterobserverVariabilityLeads2002}                             & CT   & 14.2 & 6.6  & --  & -- \\
    Sanders 2023, 7 observers~\cite{sandersUncertaintyMagneticResonance2023a}             & MRI  & 16.5 & 5.0  & 4.0 & 2.5$^{\dagger}$ \\
    De Brabandere 2012~\cite{debrabandereProstatePostimplantDosimetry2012}                    & CT   & 25.1 & 11.0 & 7.8 & 4.7$^{\dagger}$ \\
    \midrule
    \textbf{Retained margin}              &      & \textbf{10.0} & \textbf{3.0} & \textbf{4.0} & \textbf{4.0} \\
    \bottomrule
  \end{tabular}
\end{table}

For $V_{200}$ the equivalence margin value was extrapolated from the
multiplicative growth of relative dispersion between successive isodose levels. This
factor has been documented between 1.14 and 1.35 across published references and computed at 1.32 to
1.33 on the $V_{150}$-to-$V_{200}$ step in our cohort.  

Contouring variability's impact on dose-volume indices value is well documented for the prostate and for the
neighbouring organs at
risk~\cite{barghiImpactContouringVariability2013,bhardwajVariationsInterobserverContouring2008,rosewallEffectDelineationMethod2011,chicas-settEvaluationRobustnessOrganatrisk2016,chicas-settInterobserverVariabilityRectum2018},
but only studies reporting post-implant prostatic indices on a scale convertible
to ours enter Table~\ref{tabS:margins}.

\subsection{Interpretation of the two individual-agreement descriptors}

We used two descriptors to quantify the patient-level equivalence with regard to our determined margins.
The first applies the margin to each patient individually giving the fraction of the cohort that showcases equivalent DVH indices. It is the quantity the equivalence threshold regression models directly.

The second is Lin's concordance correlation coefficient~\cite{linConcordanceCorrelationCoefficient1989}, preferred to Pearson's
because the latter is blind to a systematic offset~\cite{schoberCorrelationCoefficientsAppropriate2018}. It is reported for
comparability with the concordance literature~\cite{giavarinaUnderstandingBlandAltman2015a,mcgrawFormingInferencesIntraclass1996} and because it requires no margin,
so a reader who chooses different margins $\delta$ has a descriptor of patient-level equivalence to compare with. Its interpretation must nonetheless be qualified. With a
small bias it reduces to $\rho_{c}\approx1-\tfrac{1}{2}(s_{d}/s_{b})^{2}$, where
$s_{d}$ is the standard deviation of the paired differences and $s_{b}$ the
between-patient dispersion of the index, so it measures contouring noise relative
to cohort heterogeneity rather than against any dosimetric tolerance. An index
whose values vary little across patients is penalised even when its absolute
agreement is good.

\subsection{Choice of the segmentation quality metric}

Thresholds are reported solely on the Dice coefficient. Since the choice of a segmentation quality
metric can be up to debate~\cite{lebaoEvaluatingRelationshipContouring2024,vanaalstDosebasedEvaluationDelineation2026}, eight candidate metrics, spanning overlap, distance and
volumetric families, were compared as predictors of equivalence ($\mathrm{E}=\mathbf{1}\{\lvert\Delta\rvert\le\delta\}$) within the
same framework. The Dice coefficient was the best discriminator. Because all metrics are measured on
the same patients and are therefore correlated, they were compared by a paired
bootstrap in which patients are resampled once per replicate and the difference
in area under the curve is recomputed on that same resample, so that the interval
applies directly to the difference.

\begin{figure}[pos=H]
    \centering
    \includegraphics[width=1\linewidth]{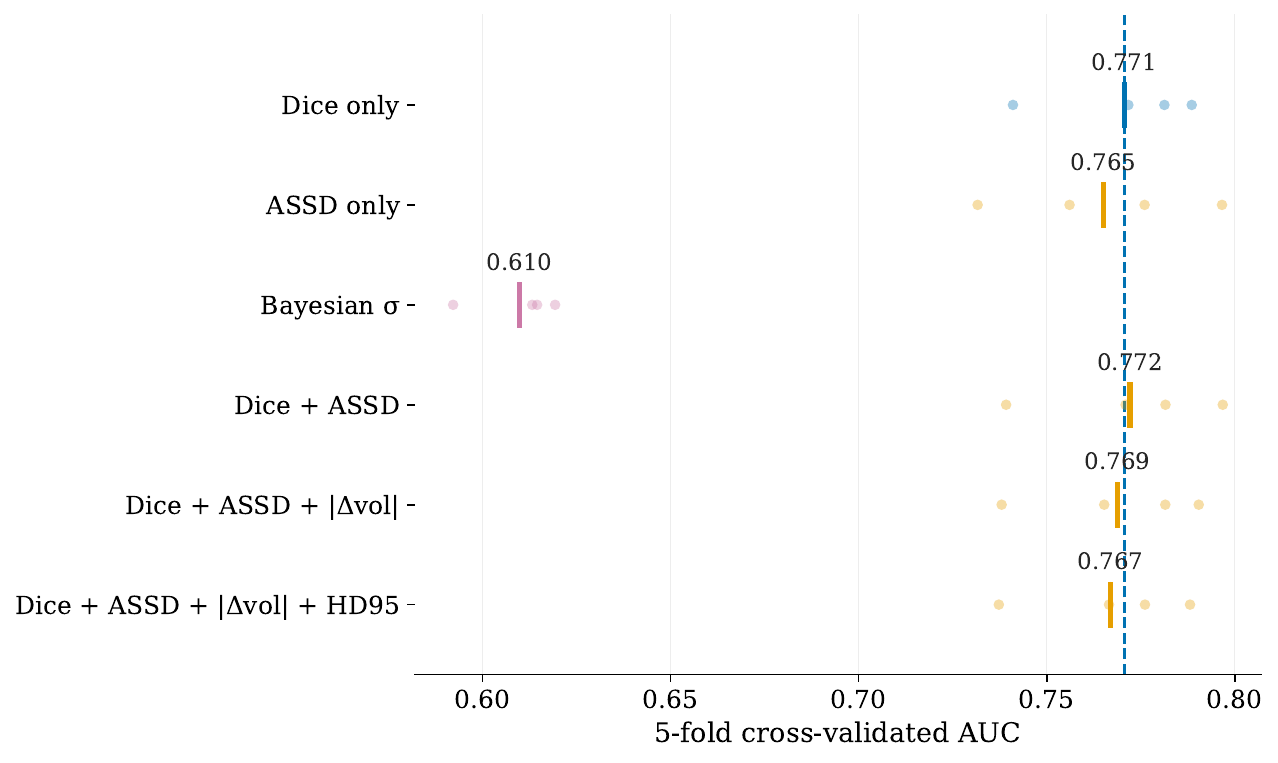}
    \caption{Cross-validated discrimination of the segmentation quality metrics as classifiers of
patient-level equivalence. Each pale marker is one index at its declared margin, for the
deterministic source, and the vertical bar is their mean. We only show the 5 best candidates, of both single and multi-metric criteria, for discrimination. We also show the Bayesian variance as an equivalence discriminator.}
    \label{fig:metric_choice}
\end{figure}

Across the 28 cells opposing the eight candidate metrics to the four
indices (4 cells compare Dice on Dice), not one paired bootstrap interval on the difference
in area under the curve excluded zero in favour of a competitor. As can be seen from Fig.~\ref{fig:metric_choice}, on cross-validated discrimination, Dice averaged 0.771 against
0.765 for its closest competitor, the average symmetric surface distance (ASSD). Combining metrics gained nothing either. The best pair, Dice with ASSD, averaged 0.772 against 0.771 for Dice alone, and
that gain of one thousandth is carried entirely by $V_{200}$. Adding the absolute volume error and
then the 95th percentile Hausdorff distance lowered the mean again, to 0.769 and
0.767. Reporting a
single metric costs no discrimination and keeps the threshold interpretable.

We also looked at the Bayesian variance, as a discriminator for equivalence. As seen in Fig.~\ref{fig:metric_choice}, it performed poorly as a discriminator, ruling out the Bayesian variance as a good metric for the segmentation quality, on which to base the logistic model of the patient-level equivalence analysis of the main text.

\subsection{Threshold estimator and the volume-adjusted models}
\label{sup:threshold_models}

An estimator determines dosimetric equivalence of an automatic segmentation by taking into account an equivalence margin $\delta$ and a segmentation quality score $s$. We used logistic regressions of the binary equivalence indicator $\mathrm{E}=\mathbf{1}\{\lvert\Delta\rvert\le\delta\}$ on $s$, 

\begin{equation}\label{eqS:logistic}
  \operatorname{logit}\Pr(\mathrm{E}=1\mid s)=\beta_{0}+\beta_{1}s,
\end{equation}
inverted at a target probability $p^{*}=0.90$ to give
\begin{equation}\label{eqS:logistic_threshold}
  s^{*}=\frac{\operatorname{logit}(p^{*})-\beta_{0}}{\beta_{1}} .
\end{equation}
It tells us how good must an automatic segmentation be for nine
cases in ten to fall inside the equivalence margin.

Volume is taken into account through three nested logistic models. Writing $v$ for the
clinically recorded prostate volume, centred at the cohort median and scaled in
units of 10~cc,
\begin{align}
  M_{1}&:\quad \operatorname{logit}\Pr(\mathrm{E}=1)=b_{0}+b_{1}s, \label{eqS:M1}\\
  M_{2}&:\quad \operatorname{logit}\Pr(\mathrm{E}=1)=b_{0}+b_{1}s+b_{2}v, \label{eqS:M2}\\
  M_{3}&:\quad \operatorname{logit}\Pr(\mathrm{E}=1)=b_{0}+b_{1}s+b_{2}v+b_{3}sv. \label{eqS:M3}
\end{align}
$M_{1}$ is Equation~\ref{eqS:logistic} rewritten with $(b_{0},b_{1})$ in place of
$(\beta_{0},\beta_{1})$, so its threshold is Equation~\ref{eqS:logistic_threshold}
and does not depend on volume. Under $M_{2}$ the threshold is displaced by a
constant amount per unit of volume, $-b_{2}/b_{1}$ per 10~cc. Under $M_{3}$ it is
no longer a number but a curve,
\begin{equation}\label{eqS:threshold_curve}
  s^{*}(v)=\frac{\operatorname{logit}(p^{*})-b_{0}-b_{2}v}{b_{1}+b_{3}v} .
\end{equation}

The three models were trained for each index and compared by likelihood ratio
tests, $M_{2}$ against $M_{1}$ for the volume main effect and $M_{3}$ against
$M_{2}$ for the interaction. $M_{3}$ is only retained when it reaches $p<0.01$, and $M_{2}$ is the fallback otherwise. $M_{1}$ is never the
reported model for the threshold curves.

The interaction test gave $p=3\times10^{-5}$ on $D_{90}$ and
$p=2\times10^{-4}$ on $V_{100}$, against $p=0.023$ on $V_{150}$ and $p=0.63$ on
$V_{200}$, so $M_{3}$ was retained on the first two and $M_{2}$ on the last two.
The volume main effect gave 0.039, 0.44, 0.083 and 0.005 respectively. The
retained model was then refitted and evaluated at the first decile, the three
quartiles and the ninth decile of the volume distribution to produce the
threshold curve of the main text. 

\section{Sensitivity of the corrected resampled \texorpdfstring{$t$}{t}-test}\label{sup:power}
 
The results of the predictive analysis point toward a lack of predictive value of the DVH panel for IPSS, when added to the clinical variables, in our cohort. This result holds only if our analysis was built, in the first place, in a way which would allow us to detect added predictive value. This section
describes the cross-validation procedure from which the per-fold differences were computed, and derives what the corrected test would be able to detect, given a significant incremental predictive value.
 
\subsection{Cross-validation training procedure}\label{sup:cv_procedure}
 
\begin{figure}[pos=H]
  \centering
  \includegraphics[width=\linewidth]{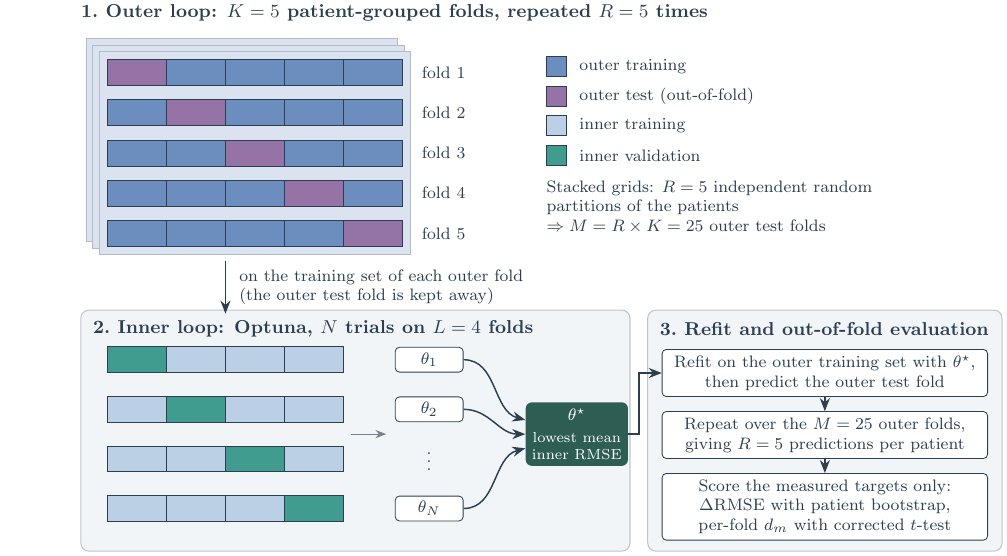}
  \caption{Repeated nested cross-validation used for the continuous target.
    Hyperparameters are tuned on the outer training set only (panel 2), and the
    model refitted with $\theta^{\star}$ predicts the outer test fold
    (panel 3). Metrics are computed on measured targets only.}
  \label{figS:nested_cv}
\end{figure}

All learners and feature blocks of the predictive analysis were trained by 5 repetitions of nested cross-validation (Figure~\ref{figS:nested_cv}). For a given horizon, the patients with real measurements and those with an extrapolated measurement were partitioned into $K=5$ outer folds, in which the patients were the sampling unit. This random partition was repeated $R=5$ times giving a total of $M=RK=25$ outer folds.
Within
each outer fold, Optuna~\cite{akibaOptunaNextgenerationHyperparameter2019}
tuned the hyperparameters on the outer training set, ranking $N$
possible configurations ($N=50$, or 40 for XGBoost) by their mean RMSE over the
$L=4$ inner folds. The best configuration $\theta^{\star}$ was then refitted on
the outer training set to predict the outer test fold.

Each patient has $R$ out-of-fold predictions. The
$\Delta\mathrm{RMSE}$ pools them over the measured targets, and its bootstrap
resamples patients with all their predictions. Within outer fold $m$, the same
targets give the per-fold difference
$d_{m}=\mathrm{RMSE}_{m}(M_{0})-\mathrm{RMSE}_{m}(M_{1})$, on
which the test below operates.
 
\subsection{Why the naive test is inadequate}

The per-fold differences $d_{1},\dots,d_{M}$ are not independent. Any two outer
training sets of a $K$-fold cross-validation procedure share $K-2$ training groups, so the $d_{m}$ are positively
correlated~\cite{NIPS2003_e82c4b19}. The naive standard error $s_{d}/\sqrt{M}$ ignores that correlation, underestimating the variance of $\bar{d}$, which inflates the
statistic and the type-I error rate. Nadeau and Bengio's correction~\cite{nadeauInferenceGeneralizationError2003} replaces it
by
\begin{equation}\label{eqS:NB}
  \mathrm{SE}=\sqrt{\left(\frac{1}{M}+\rho\right)s_{d}^{2}},
  \qquad \rho=\frac{n_{\rm test}}{n_{\rm train}}=\frac{1}{K-1},
\end{equation}
the added term $\rho$ standing for the covariance induced by the
overlap. With $K=5$ outer folds, $\rho=0.25$, and with five repetitions of the
outer partition, $M=25$ and the statistic is referred to a $t$ distribution on
$M-1=24$ degrees of freedom.

\subsection{The detectable effect is scale-free}

Write $\lambda=\bar{d}/s_{d}$, the mean per-fold difference expressed in units of
its own dispersion across folds. Substituting Equation~\ref{eqS:NB} into the
statistic, $s_{d}$ cancels:
\begin{equation}\label{eqS:scalefree}
  t_{\rm NB}=\frac{\bar{d}}{\mathrm{SE}}
  =\frac{\lambda}{\sqrt{1/M+\rho}} .
\end{equation}
With $M=25$ and $\rho=0.25$ the denominator of Equation~\ref{eqS:scalefree} is
$\sqrt{0.29}=0.5385$. A two-sided test at the 5\,\% level rejects when
$\lvert t_{\rm NB}\rvert>t_{0.975,24}=2.064$, that is when
\begin{equation}\label{eqS:lambdastar}
  \lambda \ge \lambda^{*}=2.064\times0.5385=1.11 .
\end{equation}
Power at a given $\lambda$ follows from the non-central $t$ distribution with
non-centrality parameter $\lambda/0.5385$ on 24 degrees of freedom
(Table~\ref{tabS:power}). 80\,\% power is reached at $\lambda=1.57$ and 90\,\% at $\lambda=1.82$. The rejection threshold $\lambda^{*}=1.11$ sits
near the fifty percent power point, as expected.

\begin{table}[pos=H]
  \centering
  \caption{Power of the corrected resampled $t$-test as a function of the
    standardised per-fold effect $\lambda=\bar{d}/s_{d}$, for $M=25$,
    $\rho=0.25$ and a two-sided test at the 5\,\% level on 24 degrees of freedom.}
  \label{tabS:power}
  \begin{tabular}{cc}
    \toprule
    $\lambda$ & Power \\
    \midrule
    1.00 & 0.43 \\
    1.11 & 0.51 \\
    1.32 & 0.65 \\
    1.40 & 0.70 \\
    1.57 & 0.80 \\
    1.82 & 0.90 \\
    \bottomrule
  \end{tabular}
\end{table}

\subsection{Further repetition does not buy sensitivity}

As $M\to\infty$ the standard error converges to $\sqrt{\rho}\,s_{d}$
rather than to zero. The threshold $\lambda^*$ converges to $\lambda^{*}\to1.96\times0.5=0.98$ and the 80\,\% power point converges to
$1.40$. Adding an infinite number of repetitions would
therefore move the rejection threshold from 1.11 to 0.98 and the 80\,\% point
from 1.57 to 1.40, gains of roughly one tenth. Our $M=25$ design already
operates close to the ceiling of what repeated $K$-fold cross-validation can provide. Further sensitivity would require raising $K$, which lowers $\rho$,
at the cost of smaller test folds and noisier per-fold estimates.

\section{Baseline performance and available headroom}\label{sup:headroom}
A null result on the DVH indice's predictive power is informative only if the clinical baseline $M_0$ left something to gain. This section shows that it did.

\subsection{The measurement floor and its estimation}
The IPSS survey score, being a patient-reported outcome, intrinsically carries a standard deviation $\sigma_1$. The predictive target is a difference between two scores, so the error of the difference is $\sigma_\Delta=\sigma_1\sqrt{2}$. This means that no model can have a RMSE below $\sigma_\Delta$, since at that magnitude of error the predictive gain cannot be distinguished from the questionnaire noise. We call $\sigma_\Delta$ the measurement floor and RMSE$_{M_0}-\sigma_\Delta$ the headroom of the baseline, representing the predictive gain that a better model could possibly reach. Accordingly, no predictor can exceed $R^2_{\rm max}=1-\sigma_\Delta^2/\mathrm{Var}(\Delta\mathrm{IPSS})$.

$\sigma_{1}$ depends on a reliability coefficient $r$, defined by
$\sigma_{1}=\mathrm{SD}_{\text{base}}\sqrt{1-r}$, where
$\mathrm{SD}_{\text{base}}=5.79$ points is the standard deviation of the
pre-treatment IPSS score, over the 593 patients with an available measure at our institution. A study reported a test-retest value of 0.90, in the French-Canadian adaptation of the IPSS survey, completed twice by 205 men at an average interval of 10.5 days~\cite{gregoireValidationFrenchAdaptation1996}. This study matches the
language of administration used in our work, and it describes stability between two measurements. Internal consistency was reported at 0.82 in the same study, and it describes the sampling of items within a same measurement rather than stability between measurements. The value of 0.92 reported for the original English version of the survey is given for comparison.

\begin{table}[pos=H]
  \centering
  \caption{Reliability coefficients available for the IPSS and the measurement
    floor each implies for a change score, given a pre-treatment standard
    deviation of 5.79 points.}
  \label{tabS:reliability}
  \begin{tabular}{lccc}
    \toprule
    Coefficient & $r$ & $\sigma_{1}$ & Floor $\sigma_{\Delta}$ \\
    \midrule
    Internal consistency, French-Canadian adaptation       & 0.82 & 2.46 & 3.47 \\
    Test-retest of the total index, same adaptation        & 0.90 & 1.83 & 2.59 \\
    Test-retest of the total index, original English index & 0.92 & 1.64 & 2.32 \\
    \bottomrule
  \end{tabular}
\end{table}

\subsection{Allowed headroom at every horizon}

Even under the highest floor presented in Table~\ref{tabS:reliability} (internal consistency coefficient), the baseline still allows at least 1.17 IPSS
points of predictive gain at every horizon. Under the two test-retest coefficients the headroom exceeds 2 points everywhere (Table~\ref{tabS:baseline_headroom}). The
largest shift produced by any DVH indices block is 0.15 points and the median over the twenty-four tests (four DVH blocks at six horizons) is 0.02. The predictive gains are thus at least ten times smaller than the headroom available. This headroom analysis shows that room existed for predictive gain.

\begin{table}[pos=H]
  \centering
  \caption{Out-of-fold performance of the clinical baseline $M_{0}$ (ElasticNet,
    measured targets only) and headroom above the measurement floor of the IPSS.
    The headroom interval spans the reliability scenarios not refuted by the
    data.}
  \label{tabS:baseline_headroom}
  \begin{tabular}{lcccc}
    \toprule
    Endpoint & $n$ & RMSE & $R^{2}$ & RMSE Headroom (IPSS points) \\
    \midrule
    30 days   & 354 & 6.74 & 0.142 & 3.27 -- 4.43 \\
    400 days  & 273 & 6.43 & 0.214 & 2.97 -- 4.12 \\
    730 days  & 199 & 5.60 & 0.320 & 2.14 -- 3.29 \\
    1125 days & 143 & 4.97 & 0.390 & 1.50 -- 2.66 \\
    1460 days & 187 & 4.64 & 0.386 & 1.17 -- 2.33 \\
    1865 days & 209 & 4.65 & 0.345 & 1.18 -- 2.34 \\
    \bottomrule
  \end{tabular}
\end{table}

\subsection{Clinical scale of the predictive ceiling}\label{sup:mcid_equiv}

To compare the MCID, which is a change in IPSS score for an individual patient, with $\Delta\mathrm{RMSE}$, which is a difference between two model errors, we need to convert the MCID into an equivalent $\Delta\mathrm{RMSE}$ shift. We write $e_{0}=y-\hat y_{0}$ and $e_{1}=y-\hat y_{1}$ for the
out-of-fold residuals of $M_0$ and of $M_1$, and
$d=\hat y_{1}-\hat y_{0}$ for the difference of residuals in a patient's
prediction. The residual of $M_0$ expands as
$e_{0}=(y-\hat y_{1})+(\hat y_{1}-\hat y_{0})=e_{1}+d$.
Squaring this identity and averaging over patients, with $\mathbb{E}[\cdot]$
the empirical mean over the cohort and $\mathrm{MSE}_{k}=\mathbb{E}[e_{k}^{2}]$ gives us
\begin{equation}
  \begin{aligned}
    \mathrm{MSE}_{0}=\mathbb{E}\big[e_{0}^{2}\big]
      &=\mathbb{E}\big[(e_{1}+d)^{2}\big]\\
      &=\mathbb{E}\big[e_{1}^{2}+d^{2}+2\,e_{1}d\big]\\
      &=\mathrm{MSE}_{1}+\mathbb{E}\big[d^{2}\big]+2\,\mathbb{E}\big[e_{1}d\big].
  \end{aligned}
\end{equation}
If $\mathbb{E}[e_{1}d]=0$, the gain in mean squared error equals
$c^{2}=\mathbb{E}[d^{2}]$, where $c$ is the root-mean-square correction per
patient in IPSS points. Since $\mathrm{MSE}_{k}=\mathrm{RMSE}_{k}^{2}$, the
corresponding gain in RMSE follows as a function of $c$, and setting
$c=\mathrm{MCID}=3$ gives
\begin{equation}
  \begin{aligned}
    \mathrm{RMSE}_{1}^{2}
      &=\mathrm{RMSE}_{0}^{2}-c^{2},\\
    \delta(c)=\mathrm{RMSE}_{0}-\mathrm{RMSE}_{1}
      &=\mathrm{RMSE}_{0}-\sqrt{\mathrm{RMSE}_{0}^{2}-c^{2}},\\
    \delta_{\mathrm{MCID}}=\delta(\mathrm{MCID})
      &=\mathrm{RMSE}_{0}-\sqrt{\mathrm{RMSE}_{0}^{2}-\mathrm{MCID}^{2}}.
  \end{aligned}
\end{equation}
and an upper confidence limit $b$ reads as an excluded correction
$c(b)=\sqrt{b\,(2\,\mathrm{RMSE}_{0}-b)}$ (Table~\ref{tabS:mcid_equiv}).

\begin{table}[pos=H]
  \centering
  \caption{Upper confidence limits on the RMSE gain (ElasticNet, most
    favourable block), the root-mean-square correction they exclude, and the
    RMSE gain equivalent to the MCID.}
  \label{tabS:mcid_equiv}
  \begin{tabular}{lcccc}
    \toprule
    Endpoint & $\mathrm{RMSE}_{0}$ & $b$ & $c(b)$ & $\delta_{\mathrm{MCID}}$ \\
    \midrule
    30 days   & 6.74 & 0.004 & 0.23 & 0.70  \\
    400 days  & 6.43 & 0.05 & 0.80 & 0.74  \\
    730 days  & 5.60 & 0.07 & 0.88 & 0.87  \\
    1125 days & 4.97 & 0.07 & 0.83 & 1.01  \\
    1460 days & 4.64 & 0.115 & 1.03 & 1.10  \\
    1865 days & 4.65 & 0.04 & 0.61 & 1.10  \\
    \bottomrule
  \end{tabular}
\end{table}

\section{Learner panel, control contrasts and binary transposition}\label{sup:panel}

\subsection{Learner panel and controls, continuous target}

\begin{figure}[pos=H]
  \centering
  \includegraphics[width=\linewidth]{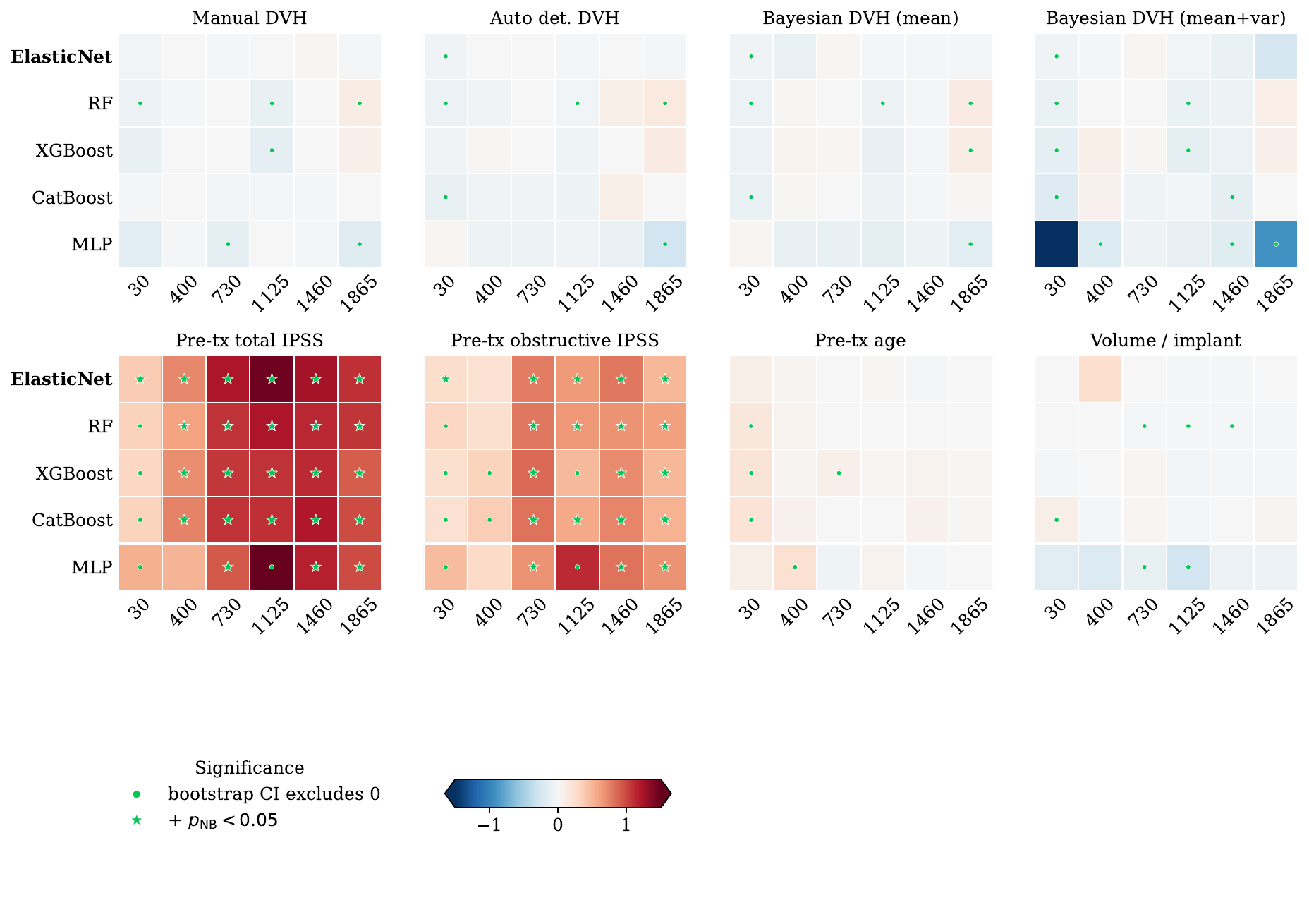}
  \caption{Incremental value of each feature block over the clinical baseline for
    the prediction of $\Delta\mathrm{IPSS}$, across the five learners and the six
    horizons. Cells report $\Delta\mathrm{RMSE}$ with the detection status under
    the declared criteria.}
  \label{figS:all_models_heatmap}
\end{figure}

The conclusion of the main text applies to all the predictive models named in the main text: ElasticNet, Random Forest, XGBoost, CatBoost, MLP
(Figure~\ref{figS:all_models_heatmap}). Over the five model types, the DVH blocks yield between zero and three bootstrap intervals
excluding zero out of six horizons, without sign consistency and never with a corrected
$p$-value below 0.05.
Ablating the pre-treatment total IPSS degrades the model's predictive power by $+0.38$ to
$+1.46$ IPSS points depending on the horizon, with six intervals out of six
excluding zero and six corrected tests out of six significant for the
ElasticNet learner. For every other model, five or six test cells satisfied the two significance criteria. Ablating the obstructive subscore alone gives $+0.23$ to $+0.79$, with significance (both criteria met) at five horizons out of six. The negative controls, age
and prostate volume, produce no detected cell at all for the ElasticNet learner
and no corrected test below 0.05 for any learner.

\subsection{Binary transposition: return to the baseline band}

We also applied the predictive analysis to a binary target, namely whether a patient remained within the MCID band at every available measurement between 1 and 5 years. Approximately half of our cohort achieved this target, which rules out class-imbalance from this analysis.
 
\begin{figure}[pos=H]
  \centering
  \includegraphics[width=\linewidth]{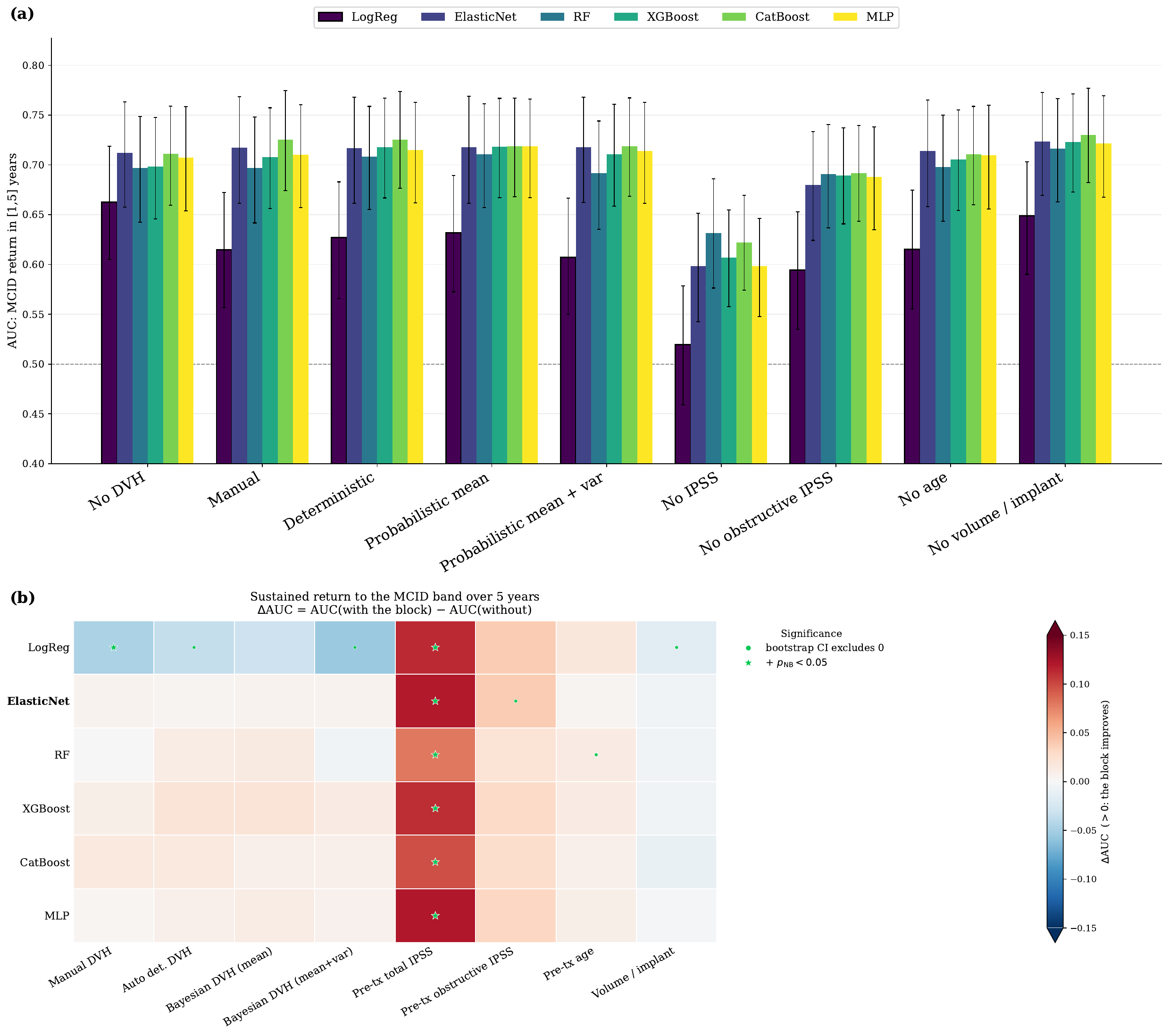}
  \caption{Binary target, return to the minimal clinically important difference
    band at every measurement between 1 and 5 years. (a) Area under the curve of
    each model. (b) Contrasts of each feature block against the clinical
    baseline, with 95\,\% bootstrap confidence intervals.}
  \label{figS:binary_auc}
\end{figure}

The ElasticNet clinical baseline $M_0$ reaches an AUC of 0.712
$[0.658,\,0.763]$ and CatBoost reaches 0.711
(Figure~\ref{figS:binary_auc}). Addition of the DVH block changes the AUC by $-0.005$ to
$+0.020$ depending on the segmentation source and the learner. No contrast AUC($M_1$)-AUC($M_0$) exhibits a
bootstrap interval excluding zero. The logistic regression model is the only model in which a DVH block reaches significance, but it does so with the
wrong sign, $-0.048$ for the manual segmentation. Ablating
the pre-treatment total IPSS costs between 0.079 and 0.120 of AUC, which is declared as significant by both criteria on all six learners. Ablating the obstructive subscore
alone costs between 0.020 and 0.038 and is detected by one bootstrap interval out
of six and by no corrected test. Age and
prostate volume move the AUC by at most 0.017. The
detection floor of the binary target therefore lies between roughly 0.04 and 0.08
of AUC, which is well above the largest shift any DVH block produces
here ($+0.020$). This is the cost of collapsing a whole trajectory into a
single target per patient at $n=340$. A DVH block moving the binary area under the curve by
less than the obstructive subscore does could not have been detected on this
target. It is also of interest to note that every continuous regression model outperforms the typical binary classifier, the logistic regression, on all feature blocks and on a binary prediction task. This points out, once more, the need to test different analysis framework when dealing with noisy data, such as patient-reported outcomes scores. 

\section{Distributional assumptions}\label{sup:distrib}

Each analysis in this study makes its own assumptions.
\begin{itemize}
  \item The TOST requires the mean paired difference, not the
    individual differences, to be approximately normal, and the patients to be
    independent.
  \item The bootstrap confidence interval on $\Delta\mathrm{RMSE}$ makes no assumption on the form of the distribution.
  \item The Nadeau-Bengio corrected $t$-test requires the mean of the 25 per-fold
    differences to be approximately normal. The correlation between folds is
    handled by the correction itself.
\end{itemize}
Normality pre-tests were not used to choose the statistical tests in this study. A pre-test alters the
distribution of the subsequent test statistic, and for the one-sample $t$
statistic it can move the actual error rate further from its nominal level~\cite{schucanyPreliminaryGoodnessofFitTests2006}.
Above about 50 observations, the central limit theorem supports the $t$
approximation and a pre-test is unnecessary~\cite{schucanyPreliminaryGoodnessofFitTests2006,rochonTestNotTest2012}.
We instead followed the recommended practice of graphical diagnostics~\cite{schucanyPreliminaryGoodnessofFitTests2006} to examine each assumption. Each conclusion relying on a normal distribution was also compared to a
distribution-free alternative.

\subsection{Two one-sided tests}

\begin{figure}[pos=H]
     \includegraphics[width=1\linewidth]{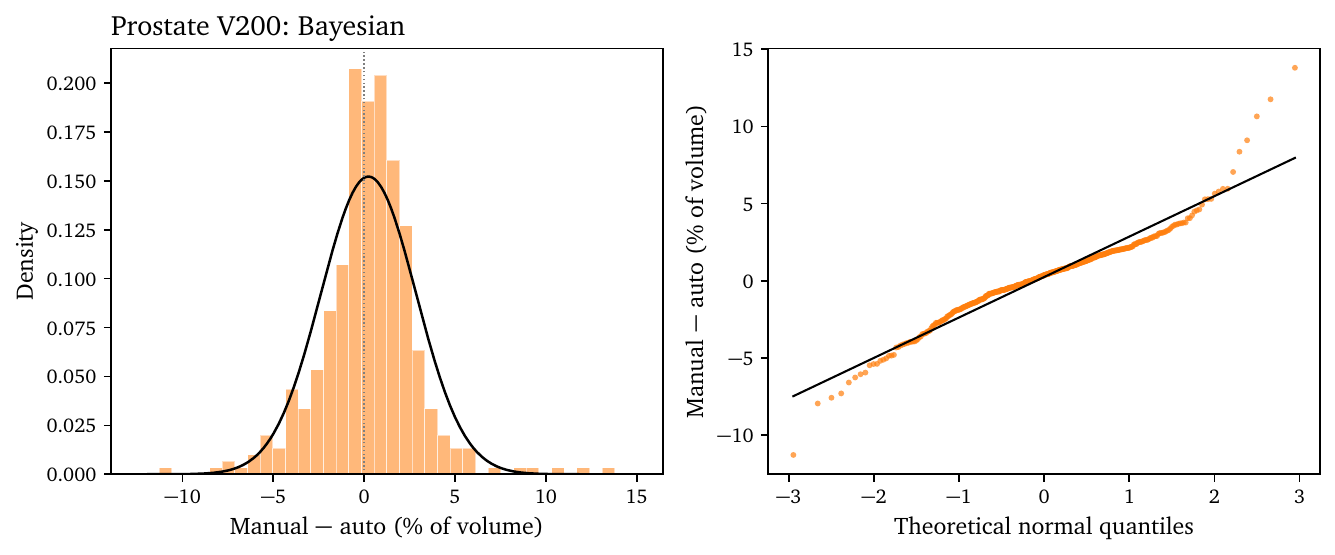}\\
    \includegraphics[width=1\linewidth]{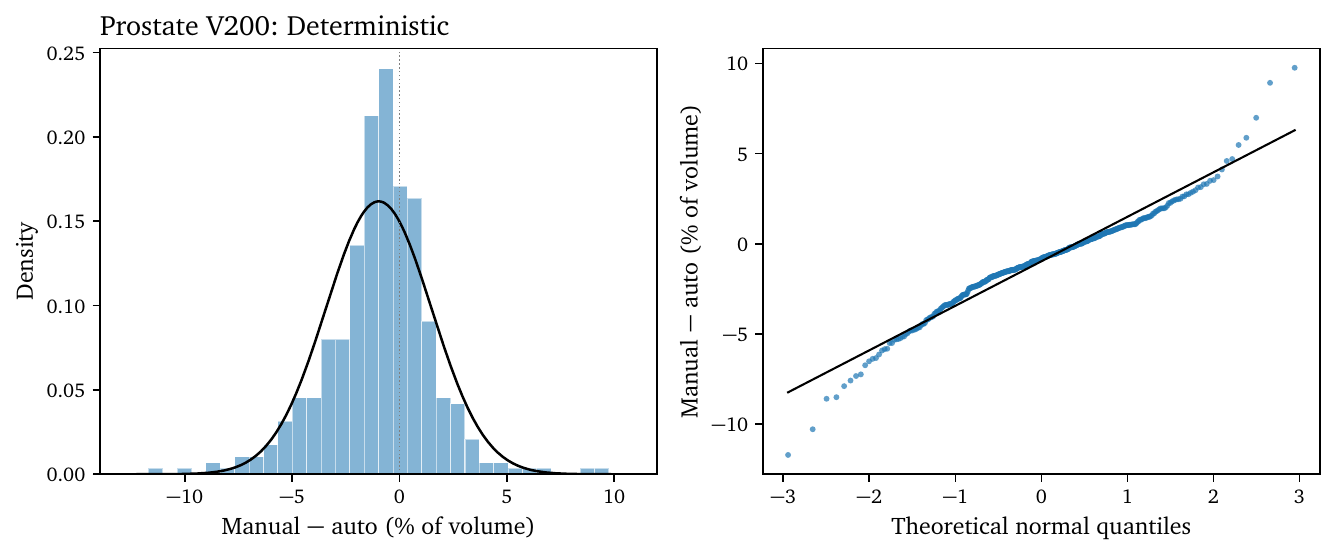}\\
    \caption{Distribution of the paired manual-automatic differences entering
  the TOST. The deterministic contours are in blue and the Bayesian ones in orange. For each comparison, the left panel shows a histogram of the fraction of the cohort in each difference bin, with a fitted normal distribution. The right panel shows the
  normal quantile-quantile plot with its least-squares reference line. Summary statistics are given in
  Table~\ref{tabS:distrib_tost}.}
    \label{figS:qq_tost}
\end{figure}

The paired differences are moderately heavy-tailed
(Table~\ref{tabS:distrib_tost}, Fig.~\ref{figS:qq_tost}), with an absolute
skewness below 0.5 for 7/8 comparison. The exception is
$V_{100}$ with the Bayesian source. The Shapiro-Wilk test rejects normality for
every comparison, which is expected at $n\approx430$. Since the TOST addresses the mean, the $t$-based 90\,\% interval was compared with a
bias-corrected and accelerated (BCa) bootstrap interval, which estimates the
distribution of the mean directly from the data with \num{10000}
resamples. With this non-parametric distribution, all eight verdicts were identical. The bounds (lower and upper) differed by at most
0.03 points, and the largest fraction of a margin consumed was 43.8\,\% with both methods.
\begin{table}[pos=H]
  \centering
  \caption{Distribution of the paired manual-automatic differences.
    All Shapiro-Wilk $p$-values are below $10^{-6}$. Dose indices in percentage
    points of prescription, volume indices in percentage points of volume.}
  \label{tabS:distrib_tost}
  \small
  \begin{tabular}{llccccc}
    \toprule
    Index & Source  & Skew. & Ex. kurt. & $t$ interval & BCa interval & $\delta$ \\
    \midrule
    $D_{90}$  & Bayesian       & $+0.15$ & 3.73  & $[-0.23,\,+1.76]$ & $[-0.21,\,+1.78]$ & 10 \\
    $D_{90}$  & Deterministic  & $-0.49$ & 2.22  & $[-4.24,\,-2.44]$ & $[-4.23,\,-2.44]$ & 10 \\
    $V_{100}$ & Bayesian       & $+1.64$ & 12.88 & $[+0.25,\,+1.04]$ & $[+0.27,\,+1.06]$ & 3  \\
    $V_{100}$ & Deterministic  & $-0.21$ & 2.00  & $[-1.30,\,-0.66]$ & $[-1.29,\,-0.66]$ & 3  \\
    $V_{150}$ & Bayesian       & $+0.43$ & 4.83  & $[-0.03,\,+0.71]$ & $[-0.03,\,+0.70]$ & 4  \\
    $V_{150}$ & Deterministic  & $-0.14$ & 2.55  & $[-1.75,\,-1.08]$ & $[-1.75,\,-1.08]$ & 4  \\
    $V_{200}$ & Bayesian       & $+0.24$ & 3.95  & $[+0.03,\,+0.45]$ & $[+0.03,\,+0.45]$ & 4  \\
    $V_{200}$ & Deterministic  & $-0.23$ & 2.67  & $[-1.17,\,-0.77]$ & $[-1.16,\,-0.77]$ & 4  \\
    \bottomrule
  \end{tabular}
\end{table}

\subsection{Bootstrap interval on \texorpdfstring{$\Delta\mathrm{RMSE}$}{Delta RMSE}}

The only requirement of the bootstrap interval is the independence of its
resampling units, which are the patients here. Patients were resampled with all their out-of-fold
predictions (for a patient having more than one prediction in a given horizon), so this requirement is met here.

\subsection{Corrected \texorpdfstring{$t$}{t}-test}

The Shapiro-Wilk test rejects normality in 38 of the 120 cells comparing a DVH block with the
baseline (5 learners with 6 horizons and 4 DVH blocks), 29 of them point toward degradation (the DVH block decreases the accuracy of prediction compared to the clinical baseline). Four of the five cells closest to the significance threshold show
distributions compatible with normality, and the fifth barely rejects normality
(skewness 0.72, Shapiro-Wilk $p=0.044$) (Table~\ref{tabS:distrib_nb}). All of these cells belong to the Random Forest learner and are not replicated in other learners, which the decision rule also
requires, so no conclusion depends on the normality assumption. Because the
per-fold differences are correlated, the Shapiro-Wilk statistics are
descriptive only.

\begin{table}[pos=H]
  \centering
  \caption{DVH cells with a positive mean per-fold difference and a corrected
    $p$-value below 0.3, all obtained with the random forest.
    $\Delta\mathrm{RMSE}$ in IPSS points with its 95\,\% bootstrap interval.}
  \label{tabS:distrib_nb}
  \small
  \setlength{\tabcolsep}{4pt}
  \begin{tabular}{llccccc}
    \toprule
    DVH source & Endpoint & $\Delta\mathrm{RMSE}$ [95\,\% CI] & Mean fold diff. & Corrected $p$ & Shapiro-Wilk $p$ & Skewness \\
    \midrule
    Deterministic       & 1865 days & $0.150$ $[0.033,\,0.268]$  & 0.152 & 0.056 & 0.044 & $+0.72$ \\
    Manual              & 1865 days & $0.112$ $[0.014,\,0.215]$  & 0.112 & 0.082 & 0.27  & $+0.39$ \\
    Bayesian            & 1865 days & $0.120$ $[0.015,\,0.230]$  & 0.120 & 0.109 & 0.43  & $+0.47$ \\
    Deterministic       & 1460 days & $0.093$ $[-0.055,\,0.221]$ & 0.093 & 0.193 & 0.83  & $+0.24$ \\
    Bayesian + variance & 1865 days & $0.096$ $[-0.018,\,0.210]$ & 0.098 & 0.228 & 0.66  & $-0.35$ \\
    \bottomrule
  \end{tabular}
\end{table}

\begin{figure}[pos=H]
    \centering
    \includegraphics[width=1\linewidth]{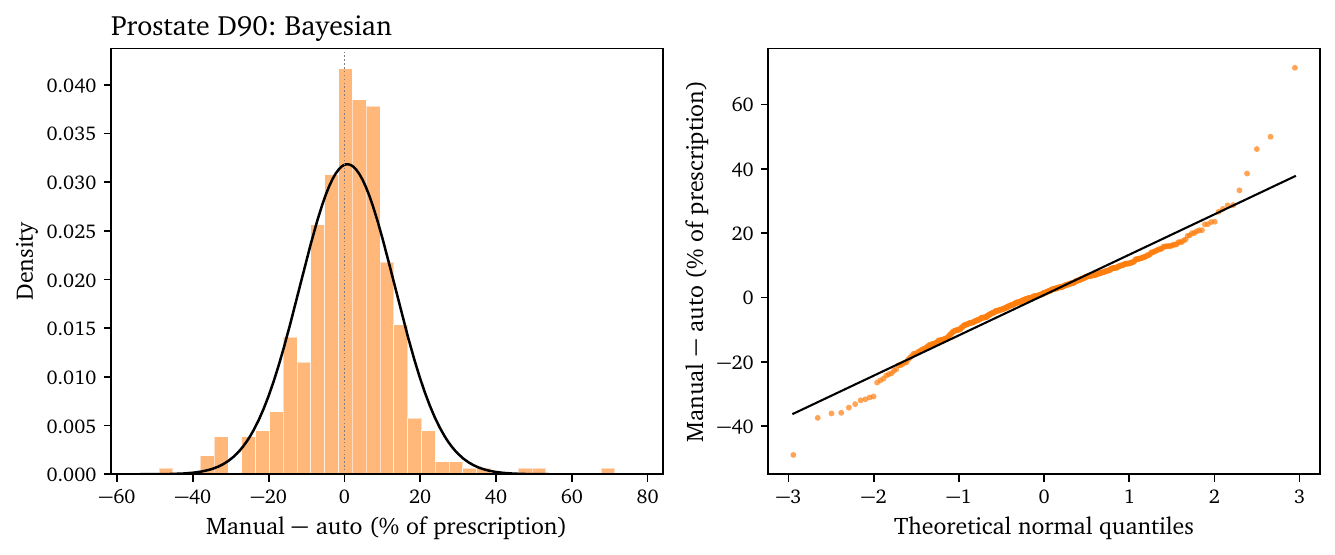}\\
    \includegraphics[width=1\linewidth]{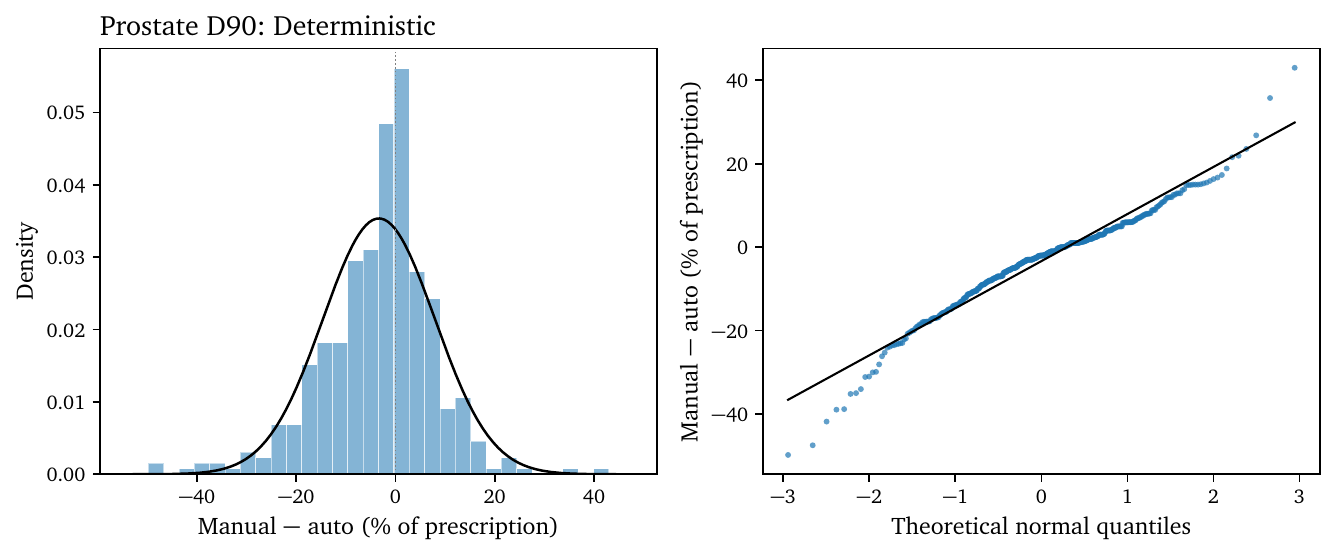}
\end{figure}
\begin{figure}[pos=H]
    \includegraphics[width=1\linewidth]{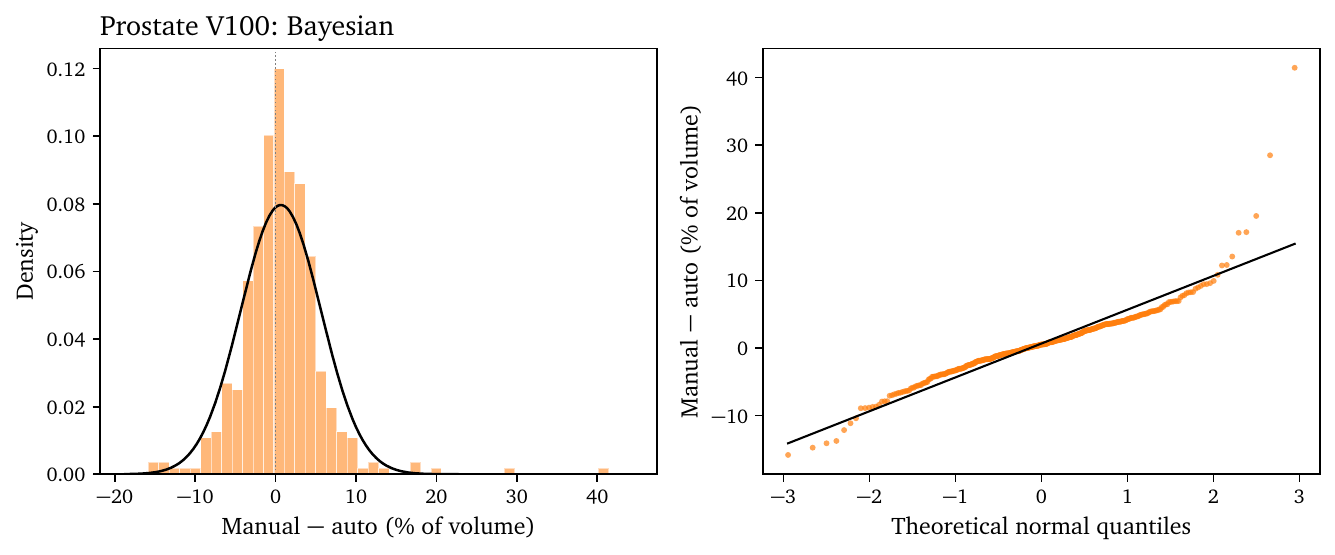}\\
    \includegraphics[width=1\linewidth]{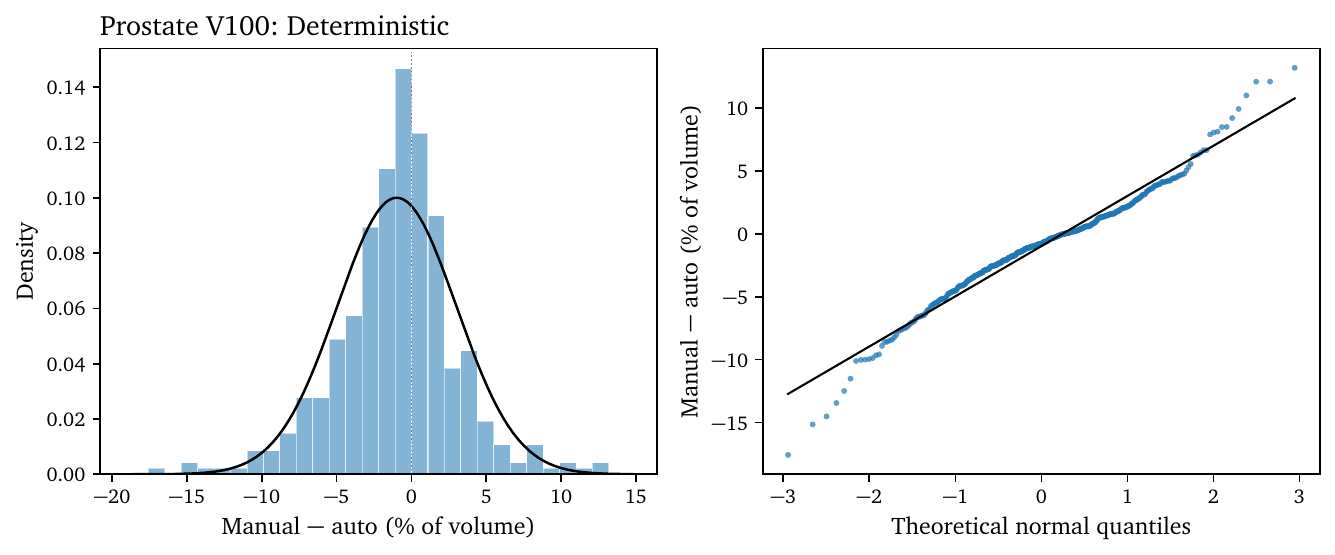}
\end{figure}
\begin{figure}[pos=H]
     \includegraphics[width=1\linewidth]{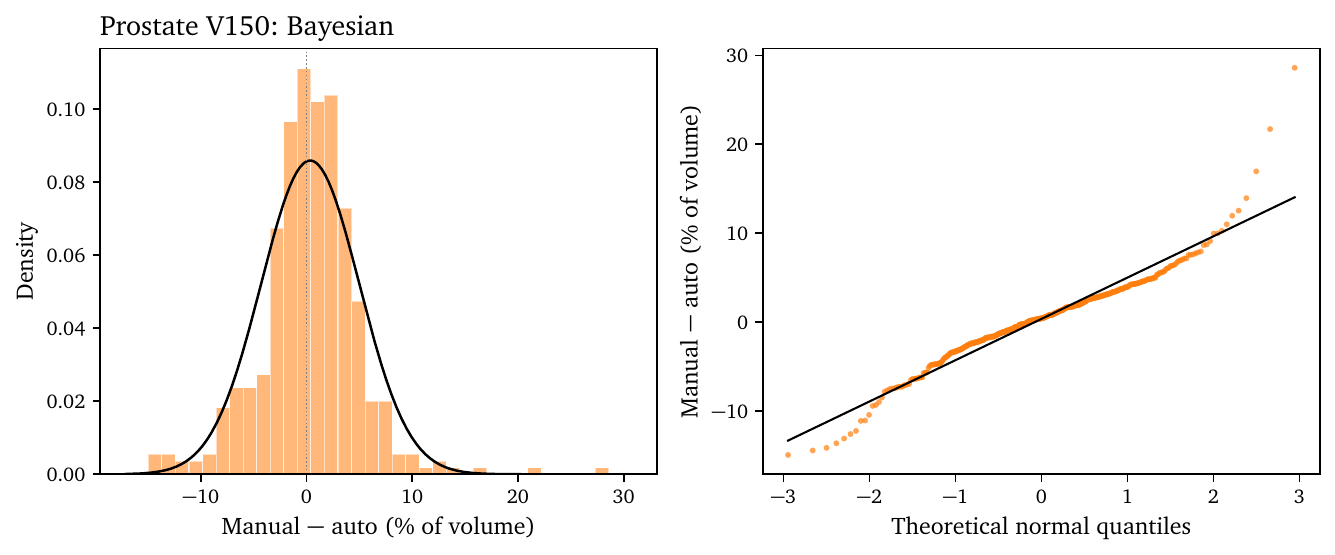}\\
    \includegraphics[width=1\linewidth]{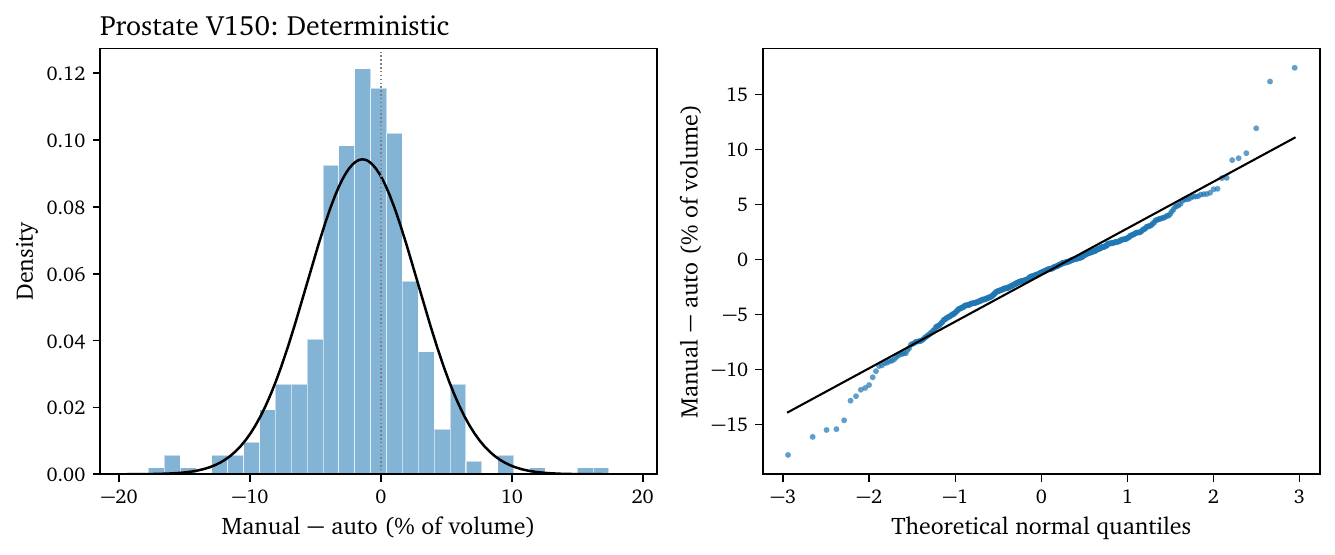}
\end{figure}

\end{document}